\documentclass[twocolumn, prd, aps, nofootinbib, superscriptaddress]{revtex4-1}
\usepackage[utf8]{inputenc}
\usepackage[english]{babel}
\usepackage{amsmath}
\usepackage{mathrsfs}
\usepackage{amssymb}
\usepackage{graphicx}
\usepackage{physics}
\usepackage{hyperref}
\usepackage{xcolor}
\usepackage{comment}

\begin{document}

\title{Overtones or higher harmonics? Prospects for testing the no-hair theorem with gravitational wave detections}

\author{Iara Ota}
\email[]{iara.ota@ufabc.edu.br}
\affiliation{Centro de Ci\^encias Naturais e Humanas, UFABC, Santo Andr\'e, SP  09210-170, Brazil}
\author{Cecilia Chirenti}
\email[]{cecilia.chirenti@ufabc.edu.br}
\affiliation{Centro de Matem\'atica, Computa\c c\~ao e Cogni\c c\~ao, UFABC, Santo Andr\'e, SP 09210-170, Brazil}
\affiliation{Department of Astronomy, University of Maryland, College Park, MD 20742-2421, USA}
\affiliation{Center for Research and Exploration in Space Science and Technology, NASA/GSFC, Greenbelt, MD 20771, USA}

\begin{abstract}
In light of the current (and future) gravitational wave detections, more sensitive tests of general relativity can be devised. Black hole spectroscopy has long been proposed as a way to test the no-hair theorem, that is, how closely an astrophysical black hole can be described by the Kerr geometry. 
We use numerical relativity simulations from the Simulating eXtreme Spacetimes project (SXS) to assess the detectability of one extra quasinormal mode in the ringdown of a binary black hole coalescence, with numbers $(\ell,m,n)$ distinct from the fundamental quadrupolar mode (2,2,0). Our approach uses the information from the complex waveform as well as from the time derivative of the phase in two different prescriptions that allow us to estimate the point at which the ringdown is best described by a single mode or by a sum of two modes.
By scaling all amplitudes to a fiducial time $t_{\rm peak}+10M$ ($t_{\rm peak}$ is the time of maximum waveform amplitude) our results for non-spinning binaries indicate that for mass ratios of 1:1 to approximately 5:1 the first overtone (2,2,1) will always have a larger excitation amplitude than the fundamental modes of the other harmonics (2,1,0), (3,3,0) and (4,4,0), making it a more promising candidate for detection. 
Even though the (2,2,1) mode damps about three times faster than the fundamental higher harmonics and its frequency is very close to that of the (2,2,0) mode, its larger excitation amplitude still guarantees a more favorable scenario for detection, as we show in a preliminary Rayleigh criterion + Fisher matrix mode resolvability analysis of a simulation with non-zero spin consistent with GW150914.
In particular, for non-spinning equal-mass binaries the ratio of the amplitude of the first overtone (2,2,1) to the fundamental mode (2,2,0) will be $\gtrsim 0.65$, whereas the corresponding ratio for the higher harmonics will be $\lesssim 0.05$. For non-spinning binaries with mass ratios larger than 5:1 we find that the modes (2,2,1), (2,1,0) and (3,3,0) should have comparable amplitude ratios in the range $0.3 - 0.4$.
The expectation that  the (2,2,1) mode should be more easily detectable than the (3,3,0) mode is confirmed with an extension of the mode resolvability analysis for non-spinning cases with larger mass ratios, keeping the mass of the final black hole compatible with GW150914.
\end{abstract}

%\keywords{}

\maketitle

\section{Introduction}
During their first two observing runs, O1 and O2, the LIGO-Virgo collaboration has detected gravitational waves (GWs) from ten binary black hole (BBH) mergers and one binary neutron star merger \cite{LIGOScientific:2018mvr}. During the current observing run (O3), approximately one BBH merger is detected every week \cite{GraceDB}. These detections represent an unparalleled feat of technological achievement and a triumph of general relativity (GR) (and of the numerical relativity (NR) simulations), with ongoing consequences for our understanding of astrophysics and fundamental physics. In particular, the current BBH mergers enable us to start performing some long-sought tests of GR (see for example \cite{Yunes:2013dva,ligo-testsGR,Barack_2019}).

In the ringdown phase after the merger, the GWs can be well approximated as a linear superposition of damped sinusoids, known as quasinormal modes (QNMs) (see \cite{Kokkotas:1999bd} for a review). Their oscillation frequencies and damping times 
depend only on the properties of the final black hole. For each harmonic mode $(\ell, m)$, the waveform can be expanded as a sum of QNMs given by
\begin{eqnarray}
\psi_{\ell m} &=& \sum_n A_{\ell m n} e^{i[ \omega_{\ell m n} (t - t_i)+ \phi_{\ell m n}]} \nonumber\\
&\equiv& \sum_n \psi_{\ell m n}, \qquad t\geq t_i ,
\label{eq:strain}
\end{eqnarray}
where $n=0$ denotes the fundamental mode and modes with $n = 1, 2,\ldots$ are known as overtones, $A_{\ell m n}$ and $\phi_{\ell m n}$ are respectively the initial amplitude and phase of each mode $\psi_{\ell m n}$,  $\omega_{\ell m n} \equiv \omega^r_{\ell m n} + i \omega^i_{\ell m n}$ is the corresponding quasinormal mode complex frequency and $t_i$ is the (unknown) starting time of the ringdown, usually expected to be after the time $t_{\rm peak}$ of maximum waveform amplitude of the quadrupolar mode $(\ell, m) = (2,2)$. 

The observation of multiple modes will enable new tests of GR. According to the no-hair theorem, the oscillation frequencies and damping times of each mode should be uniquely determined by the mass and spin of the final black hole. This multi-mode analysis of the QNMs has been termed black hole spectroscopy \cite{Dreyer-2004jan,Berti-2006mar,brito-2018oct,Baibhav-2018feb} in analogy to standard electromagnetic spectroscopy, because the QNMs form the spectrum of the black hole\footnote{Even the detection of a single mode can allow for tests of some alternative theories of gravity (see for example \cite{Cardoso_2019,McManus:2019ulj}) and/or exotic models for compact objects in general relativity (see \cite{Cardoso:2019rvt} for a review and \cite{Chirenti:2016hzd} for an example of a direct test using the GW150914 data.)}.

The first GW event, GW150914 \cite{ligo-gw150914}, showed that the fundamental quadrupolar mode - labeled as the $(2,2,0)$ mode - can be detected in the signal, as long as the remnant black hole mass is such that its oscillation frequency is in the LIGO band (approximately $50 - 500\ {\rm Hz}$) and the event is strong enough to stand out of the detector's background noise. Most of the proposals for black hole spectroscopy focus on other fundamental harmonic modes $(\ell, m,0)$ and neglect the overtones of the quadrupolar mode, $(2,2,n)$ \cite{Cotesta-2018oct,Kamaretsos:2011um,Kelly:2012nd,Shi:2019hqa,thrane-2017nov,Maselli:2019mjd}. It is well known that the overtones decay much faster than the fundamental mode \cite{Kokkotas:1999bd} and this has lead to the expectation that their contribution could be neglected.

However, this picture may be oversimplified. For example, in \cite{Buonanno:2006ui} the fundamental mode and a few overtones were fitted to the ringdown of NR simulations, and the remnant properties were consistently extracted from the best fundamental mode fit and a fit containing overtones at earlier times. Later, it was shown in \cite{London:2014cma} that the addition of one overtone should increase the estimated signal to noise ratio (SNR) of a ringdown detection. 
Additionally, it was pointed out in  \cite{Baibhav-2018feb} that the inclusion of overtones in the waveform should decrease the errors in the determination of the quadrupolar mode frequencies, mass and spin of the black hole remnant. More recently, it was shown in \cite{Giesler:2019uxc} that with the inclusion of seven overtones the gravitational waveform of a BBH merger can be described by a linear sum of QNMs starting at $t_{\rm peak}$, suggesting that the final black hole is already in its linear regime. 

The observational situation is still unclear. A recent spectroscopic analysis of the GW150914 ringdown signal \cite{carullo-2019jun} found no evidence of the presence of more than one harmonic mode in the ringdown when imposing a start time of the analysis greater than 10M after $t_{\rm peak}$ and showed that the modes (3,$-3$, 0), (3, $-2$, 0), (2, 1, 0) and (2, 2, 0) are all consistent with the remnant mass and spin found in the LIGO-Virgo analysis \cite{ligo-propsgw150914}.
On the other hand, \cite{Isi:2019aib} started to analyze the GW150914 data at $t_{\rm peak}$ and found some evidence for the (2,2,1) mode. 
More recently \cite{Bhagwat:2019dtm} suggested that an SNR larger than $\approx 30$ is necessary in order to perform black hole spectroscopy with overtones for nearly-equal mass BBH mergers.

Given the expected difficulties in detecting a second mode in the ringdown of a GW event, it will be  useful to know \emph{which mode} is most likely to be the second most relevant. Our goal is to compare the first overtone of the quadrupolar mode and the fundamental modes of the first higher harmonics to present the most promising case for a detection. One known source of ambiguity is that the excitation amplitudes of these exponentially damped modes depend on the chosen initial time for the ringdown. 

Several methods for the determination of the starting time have been proposed (see for example \cite{Dorband-2006oct,berti-2007sep,thrane-2017nov,Carullo-2018nov}), but most of them consider only the fundamental quadrupolar mode. Here we adapt existing techniques to estimate the contribution of the first overtone in the quadrupolar mode using the NR simulations from the Simulating eXtreme Spacetimes project (SXS) \cite{Boyle:2019kee,SXS-catalog}. For an event similar to GW150914 we expect $R = A_{221}/A_{220} = 0.66$ (scaled at $t_{\rm peak} + 10M$), nearly 10 times larger than the contribution of the higher harmonics.

This paper is organized as follows: in Section \ref{sec:fundamental} we fit a single quasinormal mode to the ringdown to determine the parameters of the remnant black hole. 
In Section \ref{sec:overtones} we compute for one representative case the initial time from which the ringdown of the quadrupolar mode is well described by the fundamental mode and the first overtone, the initial amplitude ratio of the two modes, and we estimate the detectability and resolvability of the two modes in subsection \ref{sec:resolve}.  
In Section \ref{sec:mass_ratio} we generalize our results for increasing mass ratios  and calculate the minimum ringdown SNR for resolvability of a second mode, taking into account the inclination angle of the binary. We find that the detection of the $(2,2,1)$ mode will always be favored over other harmonics. We present our closing remarks in Section \ref{sec:conclusions}.

\section{Determination of the fundamental QNM}
\label{sec:fundamental}

Throughout this section we will use the NR waveform SXS:BBH:0305, 
which is consistent with the GW150914 signal \cite{ligo-gw150914,ligo-testsGR}.
In the SXS simulations, the remnant mass $M_f$ and dimensionless spin $a$ are extracted from the apparent horizon \cite{SXS-catalog} and, for this simulation, they are given in units of the total mass of the binary $M = M_1 + M_2$ as $M_f = 0.9520 M$ and $a = 0.6589 M$. 

If the ringdown waveform $\psi_{\ell m}$ (\ref{eq:strain}) is fully described by the fundamental mode, that is, 
$\psi_{\ell m} = \psi_{\ell m 0}$, then the time derivative $\dot{\theta}_{\ell m}$ of the complex phase defined as
\begin{equation}
\theta_{\ell m} \equiv \arctan\left[\frac{\Im(\psi_{\ell m})}{\Re(\psi_{\ell m})}\right],
\end{equation}
will be equal to the fundamental mode frequency of oscillation $\omega^r_{\ell m 0}$. However, the waveform $\psi_{\ell m}$ has overtone contributions, as given by eq.~(\ref{eq:strain}), and $\dot{\theta}_{\ell m}$ should not be simply constant and equal to $\omega^r_{\ell m 0}$. This can be seen for example in the recent work \cite{Ferguson-2019}, where $\dot{\theta}_{22}$ was computed  
in order to find a fitting formula for the final spin and it was found that $\dot{\theta}_{22}$ is not constant in the interval $t - t_\mathrm{peak}  \in [-20, 20]M$ (see Fig. 1 of \cite{Ferguson-2019}). 

Nevertheless, the overtones decay much faster than the fundamental mode. In the particular case of the quadrupolar mode of the black hole remnant produced in the simulation SXS:BBH:0305, 
the damping times of the fundamental mode and the first overtone are $\tau_{220} = 1/\omega^i_{220} = 11.8M$ and $\tau_{221} = 1/\omega^i_{221} = 3.9M$, respectively (see Table \ref{tabelafreqn}). Therefore, if we assume that all modes are excited simultaneously, after some time $t_{n=0}$ the contributions of all overtones will be negligible with respect to the fundamental mode and 
$\dot{\theta}_{\ell m}(t \geq t_{n=0}) \approx \omega_{\ell m 0}^r$. 

\begin{figure}[htb!]
	\centering
	\includegraphics[scale = 0.3]{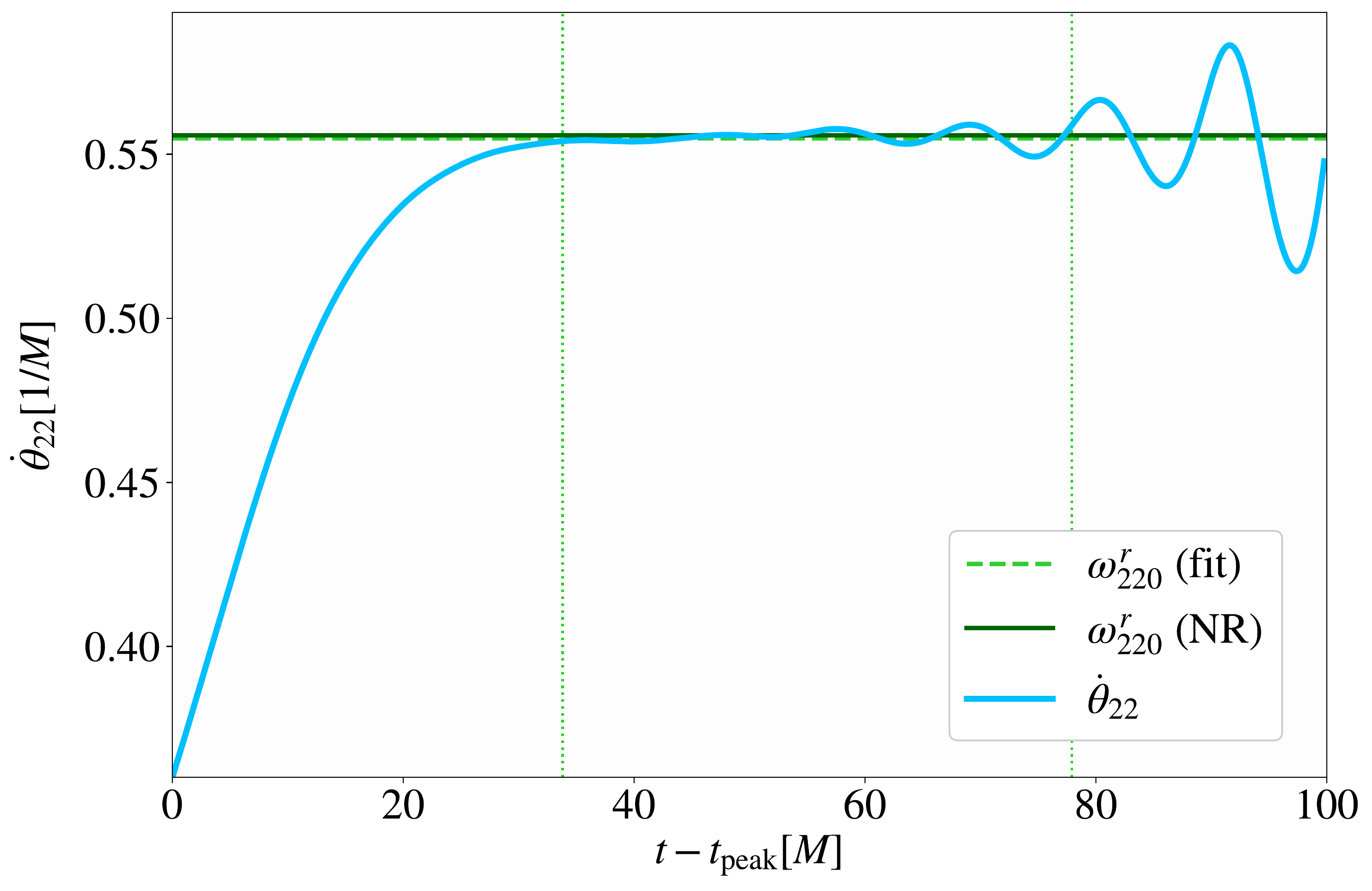}
	\includegraphics[scale = 0.3]{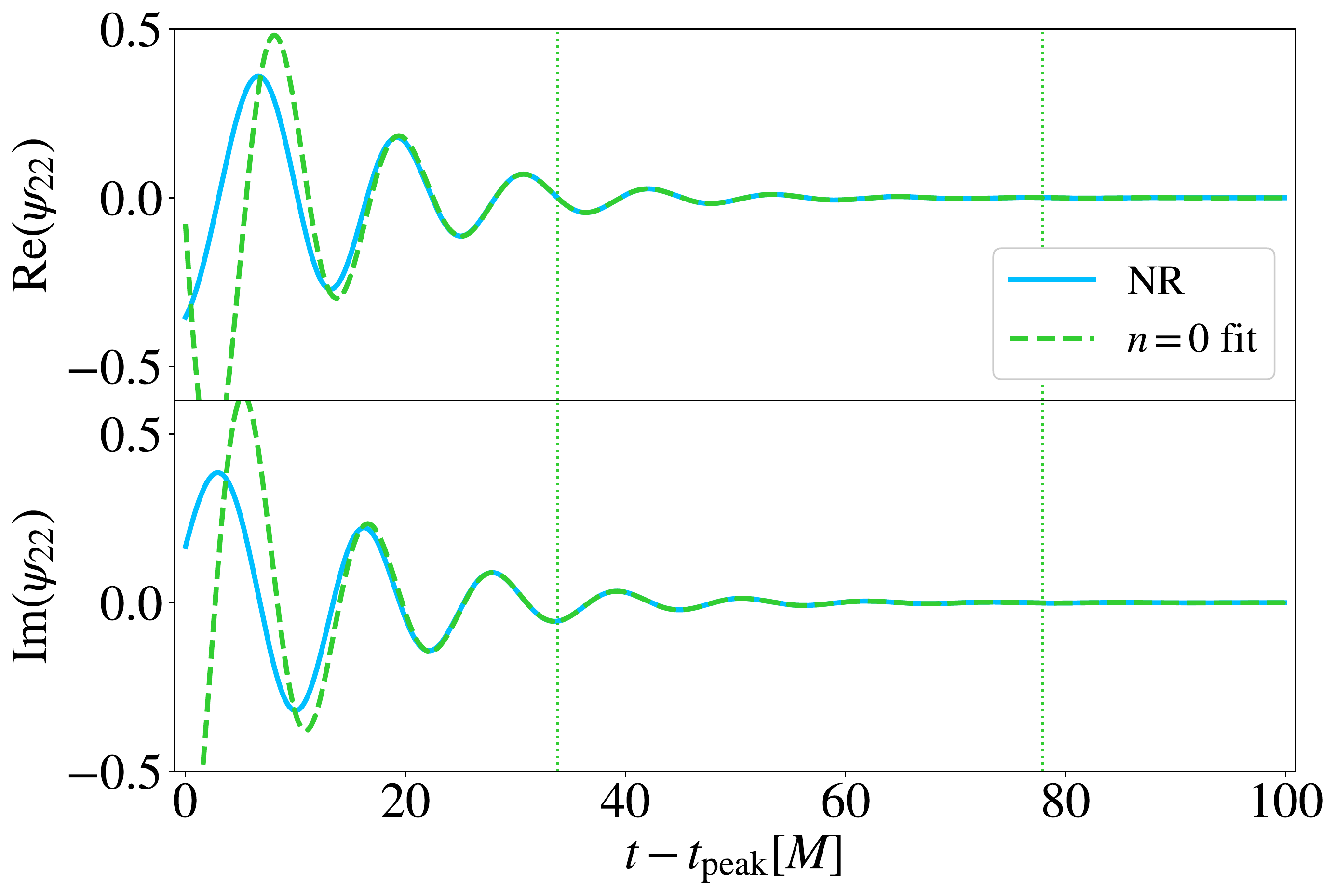}
	\caption{\emph{Top:} Time derivative of the waveform phase $\theta_{22}$ in the ringdown of the binary black hole simulation SXS:BBH:0305. The solid horizontal line indicates the value of $\omega_{220}^r$ obtained from the remnant parameters quoted in the NR simulation, while the (almost coincident) dashed line shows the value obtained by fitting the fundamental mode to the complex waveform (shown in the bottom plot) in the interval indicated by the dotted vertical lines. The oscillations at $t-t_\mathrm{peak} > 75 M$ represent late time numerical errors, where $t_\mathrm{peak}$ is the maximum waveform amplitude of the quadrupolar mode $(\ell, m) = (2,2)$. 
	\emph{Bottom:} The real and imaginary parts of the complex waveform (solid line) and the fundamental mode fit (dashed line). }
	\label{arctan_peak0}
\end{figure}

Figure \ref{arctan_peak0}  shows in the upper plot
$\dot{\theta}_{22}$ in the ringdown of SXS:BBH:0305. For $30 M \lesssim t-t_\mathrm{peak} \lesssim 75 M$ we can see that $\dot{\theta}_{22}$ approaches a constant value (with fractional variation less than approximately 1\%), after the contributions of overtones or nonlinear effects from the merger have been damped away and before numerical errors introduce larger variations at late times. In this approximate interval we have fitted the fundamental mode strain
$$
\psi_{220} = A_{0} e^{-\omega_{220}^i t}\left[\cos(\omega_{220}^r t -\phi_0) + i \sin(\omega_{220}^r t -\phi_0)\right]
$$
to the complex NR waveform shown in the lower plot, where $A_0$, $\phi_0$, $\omega_{220}^r$ and $\omega_{220}^i$ are free parameters in the fit. The frequencies obtained in this fit are $M$$\omega_{220}^r = 0.5549$ and $M$$\omega_{220}^i = 0.0848$, and the mass and spin of the final black hole computed from these frequencies \cite{Berti-2006mar,Berti-ringdown} are $M_f = 0.9553 M$ and $a_f = 0.6632 M$. This means a correction of approximately 0.3\% and 0.7\% in the quoted mass and spin of the black hole remnant, respectively.
The dashed horizontal line in the upper $\dot{\theta}_{22}$ plot presents our best fit for $\omega_{220}^r$, which differs by approximately 0.2\% from the value obtained with linear perturbation theory \cite{Berti-2006mar,Berti:2006wq,Berti-ringdown} and the quoted remnant mass and spin for this simulation, shown with the solid horizontal line. 

\begin{figure*}[htb!]
	\centering
	\includegraphics[scale = 0.2]{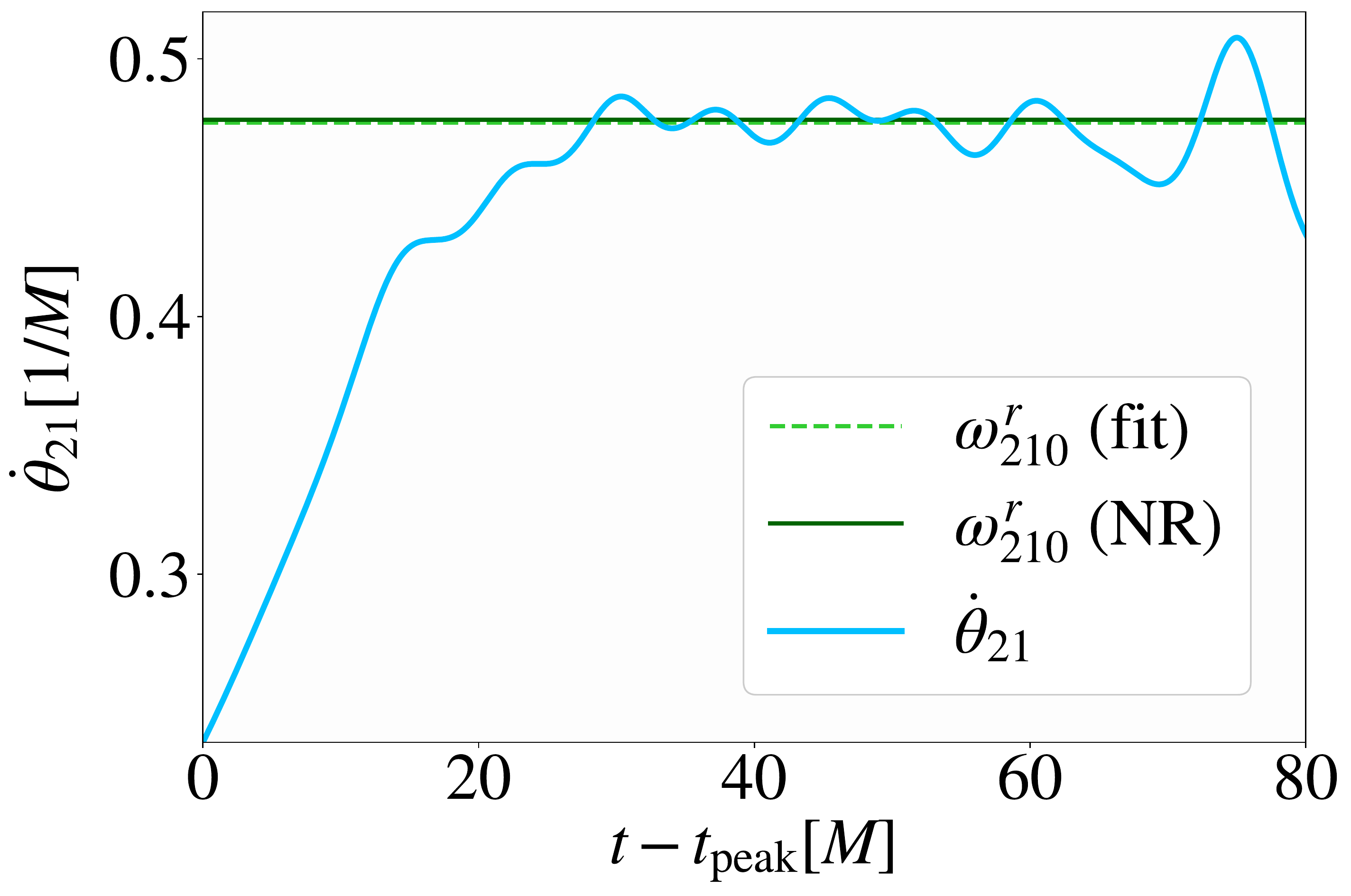}
	\includegraphics[scale = 0.2]{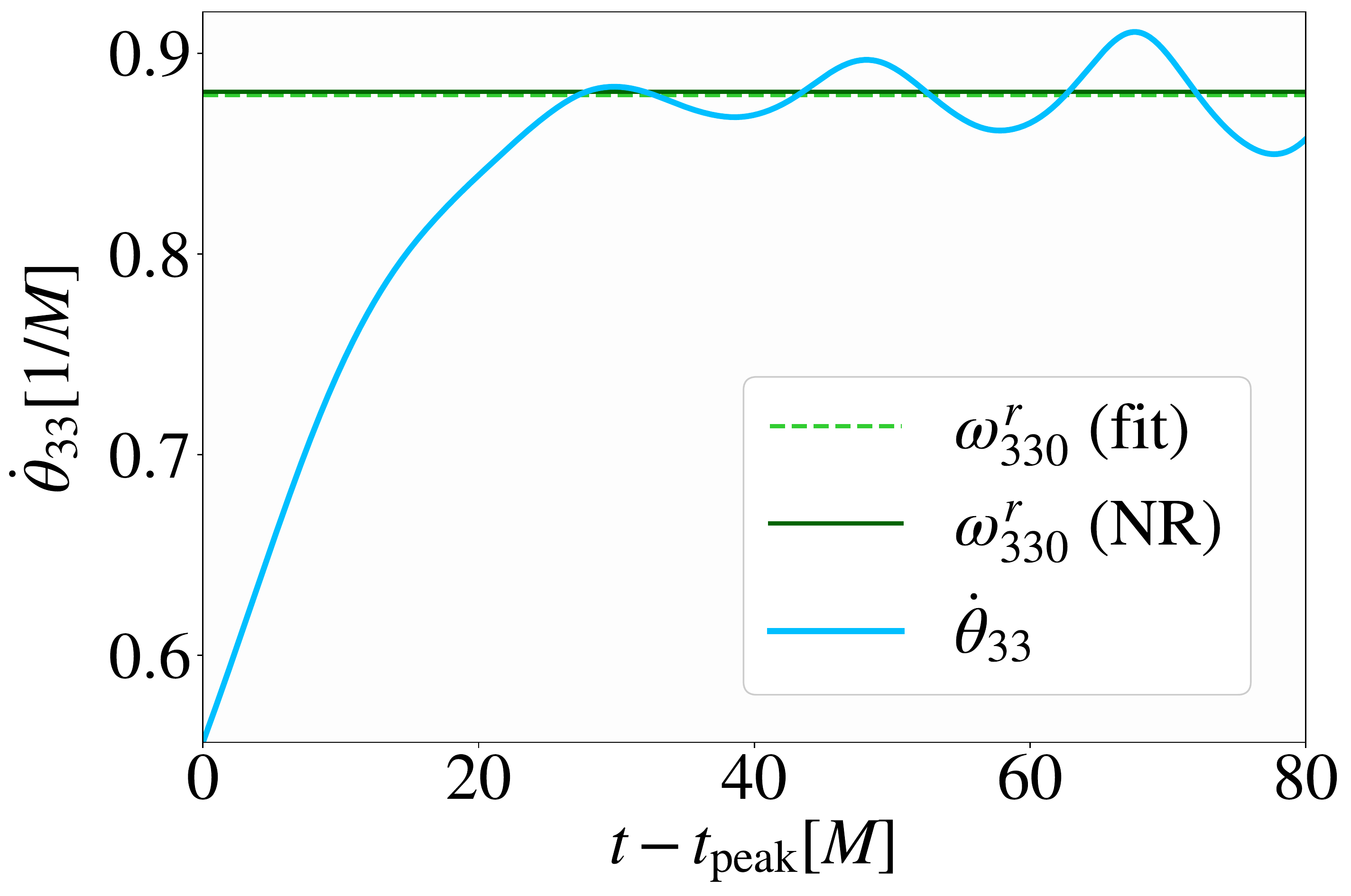}
	\includegraphics[scale = 0.2]{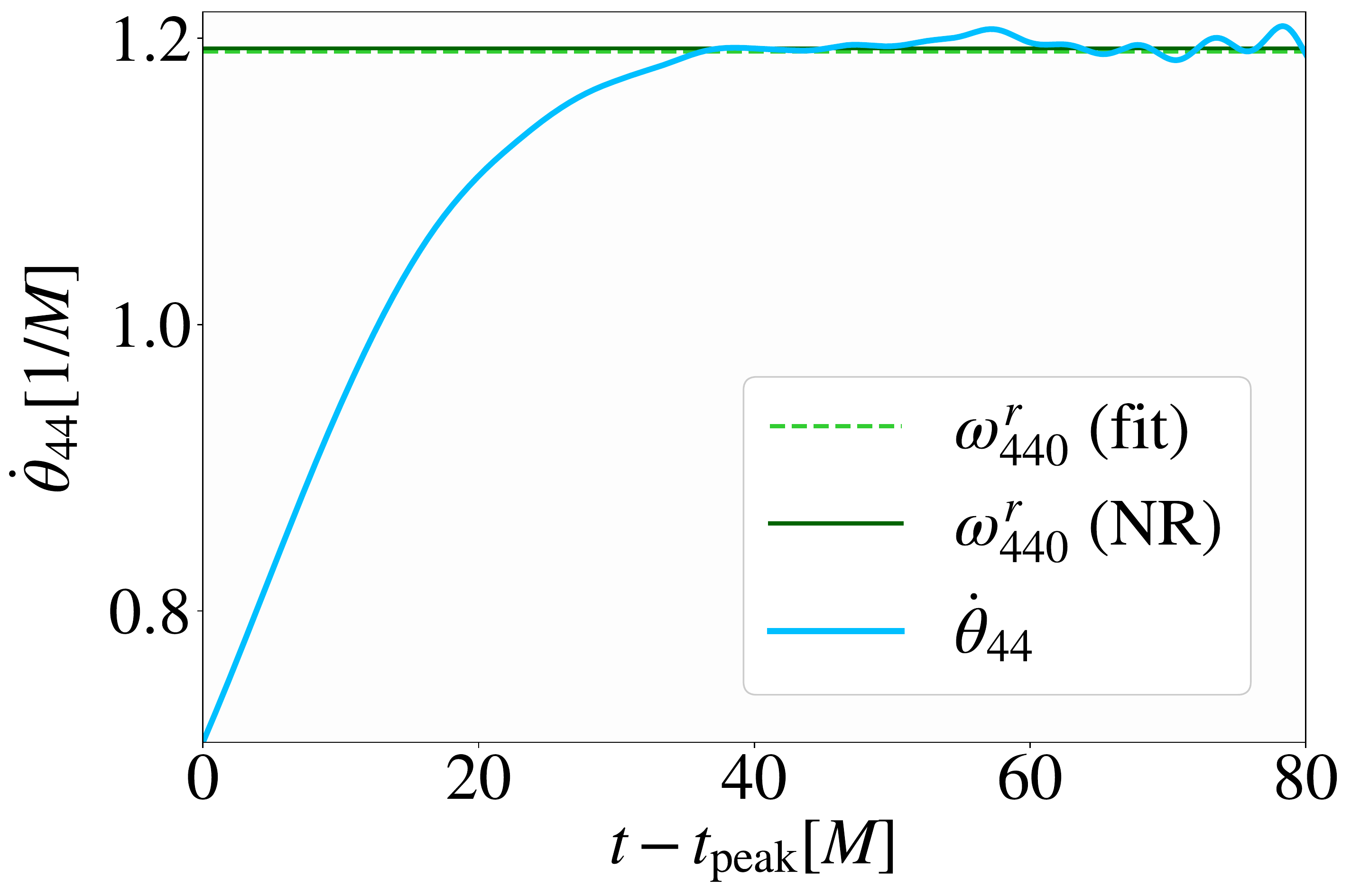}
	\caption{Same as the upper plot of Figure \ref{arctan_peak0}, but for the time derivative of the waveform phases $\theta_{21}$ (left), $\theta_{33}$ (center) and $\theta_{44}$ (right). Again we see that there is very good agreement (less than 0.2\% difference) between the values of $\omega_{\ell m 0}^r$ obtained from the remnant parameters quoted in the NR simulation and our fitted values. $\dot{\theta}_{\ell m}$ approaches $\omega_{\ell m 0}^r$ after the overtone contributions and non-linearities have decayed (and before the late time numerical errors become too relevant).}
	\label{arctan_other}
\end{figure*}

We have also looked into the other harmonics (2,1), (3,3) and (4,4). Figure \ref{arctan_other} reproduces the analysis presented in Figure \ref{arctan_peak0}, and again our results for frequencies $\omega_{210}^r$, $\omega_{330}^r$ and $\omega_{440}^r$ agree with the corresponding values obtained with the previously quoted parameters within less than approximately 0.2\%. These $\dot{\theta}_{\ell m}$ also rise from lower values towards $\omega_{\ell m 0}^r$ when the overtones and possible non-linear behavior have been damped. Numerical errors become more relevant at later times.  

\begin{table}[htb!]
\centering
\caption{Fundamental and first overtone quasinormal mode frequencies of the quadrupolar $(\ell,m) = (2,2)$ mode, obtained with linear perturbation theory \cite{Berti-2006mar,Berti-ringdown}, for a black hole remnant with parameters $M_f = 0.9553M$ and $a = 0.6632M$, inferred from SXS:BBH:0305.}
\begin{ruledtabular}
\begin{tabular}{c c c}
$n$ & $M$$\omega^r_{22n}$ & $M$$\omega^i_{22n}$ \\ \hline
0 & 0.5549 & 0.0848 \\
1 & 0.5427 & 0.2564 
\end{tabular}
\end{ruledtabular}
\label{tabelafreqn}
\end{table}

\section{Improved ringdown: fundamental mode + first overtone}
\label{sec:overtones}
It is clear from Figures \ref{arctan_peak0} and \ref{arctan_other} that the waveform immediately after the amplitude peak is not fully described by the fundamental mode, as in that case we would have a constant $\dot{\theta}_{\ell m} = \omega_{\ell m 0}^r$. The observed variation can be associated with a non-linear behavior near the peak or non-negligible contributions of overtones. The presence of the overtones can be made evident by subtracting the lowest order modes from the signal (see Fig. 7 in \cite{Chirenti_2007} for an example of this approach). In \cite{Giesler:2019uxc} it was suggested that, by including seven overtones in the model, the linear behavior of the waveform starts at the peak of the amplitude (or even before the peak). 

However, given the expected difficulties in observing and identifying other modes besides the fundamental quadrupolar mode in the gravitational wave data \cite{ligo-testsGR,carullo-2019jun,Isi:2019aib}, here we will only consider the contribution of the first overtone in the signal. We will do this by fitting to the numerical data functions containing the fundamental mode and the first overtone with frequencies given by Table \ref{tabelafreqn} and we will determine the interval of the waveform which is well described \emph{by the fundamental mode and the first overtone} \footnote{It is important to notice that the initial times we obtained do not necessarily represent the beginning of the post-merger linear regime (i.e., the ringdown) as non-negligible contributions of higher overtones $(n \ge 2)$ in the waveform are not taken into account.}.

To determine the initial time of this interval, we will use two methods. In the first method we perform a non-linear fit to the numerical waveform $\psi_{22}$ of the 4-parameter function
\begin{align}
    & \psi_{22}(t) = 
    A_{220} e^{-\omega_{220}^i t}\left[\cos(\omega_{220}^r t -\phi_{220}) + \right.\nonumber\\
    &\left.+ i \sin(\omega_{220}^r t -\phi_{220})\right] \nonumber\\ 
    &+ A_{221} e^{-\omega_{221}^i t}\left[\cos(\omega_{221}^r t -\phi_{221}) + \right.\nonumber\\
    &\left.+ i \sin(\omega_{221}^r t -\phi_{221})\right], 
    \label{eq:psi}
\end{align}
where $\omega^{(r,i)}_{22n}$ are given by Table \ref{tabelafreqn}, and the fitting parameters are the initial amplitudes $A_{22n}$  and the initial phases $\phi_{22n}$ of each mode $(n = 0,1)$. In our second method we do a non-linear fit to the numerical $\dot{\theta}_{22}$  of the 2-parameter function
\begin{align}
    \dot{\theta}_{22}(t) &= \left\{ \omega_{220}^r + R^2 e^{2(\omega_{220}^i - \omega_{221}^i)t} \omega_{221}^r + R e^{(\omega_{220}^i - \omega_{221}^i)t}\right.\nonumber\\
    &\times\left.\left[\left(\omega_{220}^r + \omega_{221}^r\right)\cos((\omega_{220}^r-\omega_{221}^r)t -\phi) \right.\right. \nonumber\\
    &\left.\left. +\, (\omega_{221}^i - \omega_{220}^i)\sin((\omega_{220}^r-\omega_{221}^r)t -\phi)\right]\vphantom{ R^2 e^{2(\omega_{220}^i - \omega_{221}^i)t}}\right\}\nonumber\\
    &\times\left[2 R e^{(\omega_{220}^i - \omega_{221}^i)t}\cos((\omega_{220}^r-\omega_{221}^r)t -\phi)\right. \nonumber\\
    &\left.  + R^2 e^{2 (\omega_{220}^i- \omega_{221}^i) t} + 1 \right]^{-1},
    \label{eq:atan}
\end{align}
where the fitting parameters are the ratio of the initial amplitudes $R \equiv A_{221}/A_{220}$ and the phase difference between the modes $\phi \equiv \phi_0 - \phi_1$. Since $\omega_{221}^i > \omega_{220}^i$, we have that $e^{(\omega_{220}^i - \omega_{221}^i)t} \rightarrow 0$ and $\dot{\theta}_{22} \rightarrow \omega_{220}^r$ as $t\rightarrow \infty$.

The initial time $t_0$ for the fits (\ref{eq:psi}) and (\ref{eq:atan}) is not treated as a fitting parameter. We  select the best initial time $t_0$ by minimizing the mismatch $\mathcal{M}$ between the NR data $f_{\mathrm{NR}}$ and the fitted function $f_{\mathrm{fit}}$, defined as 
\begin{equation}
\mathcal{M} = 1 - \frac{\langle f_{\mathrm{NR}}, f_{\mathrm{fit}} \rangle}{\sqrt{\langle f_{\mathrm{NR}}, f_{\mathrm{NR}} \rangle\langle f_{\mathrm{fit}}, f_{\mathrm{fit}} \rangle}}.
\label{eq:mismatch}	
\end{equation}
where $f$ represents either the waveform $\psi_{22}$ or the phase derivative $\dot{\theta}_{22}$. The mismatch $\mathcal{M}$ is a function of the initial time $t_0$, as \emph{the inner products in the right-hand side are computed starting at each $t_0$}. This procedure is similar to the one used in \cite{Giesler:2019uxc}. Other approaches suggested in the literature for finding the initial time of the ringdown minimize the residuals of the fit of the fundamental mode, see for example \cite{Dorband-2006oct,berti-2007sep,thrane-2017nov}.

Following \cite{8marcel-grossman}, the inner product can be defined in the usual way:
\begin{equation}
\langle f_1, f_2 \rangle_{\rm standard} \equiv \abs{\int_{t_0} f_1^* f_2 dt},
\label{eq:inner1}
\end{equation}
where the star denotes the complex conjugate. However, QNMs are not orthogonal and complete  with respect to the inner product defined above, which presents a problem for computing how much energy is contained in each mode. To circumvent this problem, Nollert \cite{8marcel-grossman,Nollert_1999} suggested an energy-oriented inner product defined as
\begin{equation}
\langle f_1, f_2 \rangle_{\rm energy} \equiv \abs{\int_{t_0} (\dot{f}_1)^* \dot{f_2} dt},
\label{eq:inner2}
\end{equation}
where the dot denotes the time derivative as before. We will use both of the inner product definitions (\ref{eq:inner1}) and (\ref{eq:inner2}) in our calculation of the mismatch (\ref{eq:mismatch}).  The mismatch will be calculated for the fits (\ref{eq:psi}) and (\ref{eq:atan}), giving four estimates for the time $t_0$, as we will see below.

\begin{figure}[htb!]
	\centering
	\includegraphics[scale = 0.32]{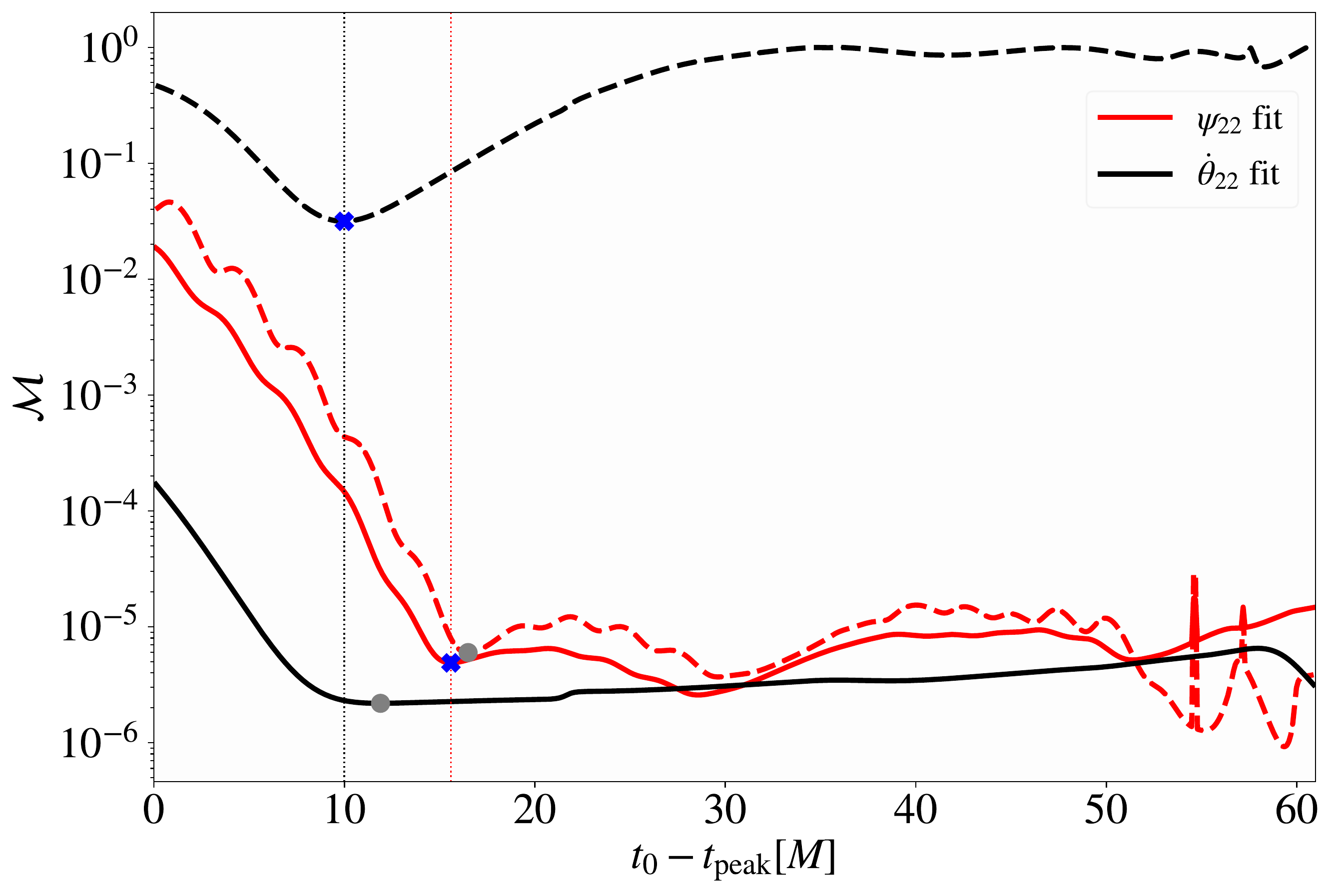}
	\caption{Mismatch (\ref{eq:mismatch}) between simulation SXS:BBH:0305 and a ringdown model with two modes (fundamental quadrupole + first overtone) as a function of time. The mismatch is calculated for the complex waveform $\psi_{22}$ (\ref{eq:psi}) in red and for the phase derivative $\dot{\theta}_{22}$ (\ref{eq:atan}) in black. Solid (dashed) lines indicate that the inner product is calculated using eq.~(\ref{eq:inner1}) (eq.~(\ref{eq:inner2})). The points show the chosen initial time $t_0$ for each case. The round gray points will not be considered in our analysis and the blue crosses are the chosen initial times for each function: $\dot{\theta}_{22}$ (method I) and $\psi_{22}$ (method II) (see text).}
	\label{fig:mismatch}
\end{figure}

Again, here we will not determine the initial time of the ringdown stage but the initial time $t_0$ at which the waveform is well described by the sum of the fundamental mode and the first overtone.
Figure \ref{fig:mismatch} shows the mismatch of the simulation SXS:BBH:0305 as a function of the initial time for the phase derivative $\dot{\theta}_{22}$  (black) and for the waveform $\psi_{22}$ (red). Solid (dashed) lines indicate that the inner product was calculated with eq.~(\ref{eq:inner1}) (eq.~(\ref{eq:inner2})). 
We choose the initial time $t_0$ as the first minimum of the mismatch $\mathcal{M}$, ignoring the local minima in the oscillations when $\mathcal{M}$ is still decreasing. The highlighted points (blue crosses and gray dots) in Figure \ref{fig:mismatch} show the initial time for each of the four calculations.

The initial time $t_0$ depends on the method (choice of inner product and fitting function).
The dashed black curve shows a clear minimum at about $t_0-t_{peak} = 10 M$ (marked with a blue cross), and the largest values for the mismatch (because $\ddot{\theta}_{22} \approx 0$ for $\dot{\theta}_{22} \approx \omega^r_{220}$). The solid black curve shows that the mismatch stops decreasing at approximately that time, but the actual first minimum is marked by the gray dot and it is typically less precisely determined because of the flattening of the curve. The dashed and solid red curves show very similar behaviors, with a minimum close to $t_0 - t_\mathrm{peak} = 15.6M$ (blue cross) but the dashed curve presents more oscillations and local minima that make the determination of $t_0$ more uncertain (gray dot). 

With these general considerations, which are typical of all simulations we have analyzed, we have chosen to keep the estimates for $t_0$ given by only two methods, which look for: 
\begin{enumerate} 
\item[I] the minimum of the mismatch of $\dot{\theta}_{22}$ computed with the energy-oriented inner product given by eq.~(\ref{eq:inner2}) 
\item[II] the first minimum of the mismatch of $\psi_{22}$ computed with the standard inner product given by eq.~(\ref{eq:inner1}),
\end{enumerate}
which are shown in Figure \ref{fig:mismatch} with blue crosses on the dashed black and solid red curves, respectively. 

Our results are compatible with the recent multimode analysis of the ringdown phase of the GW150914 detection \cite{carullo-2019jun}, where the initial time is defined as the time in which the fundamental mode (of both $\ell = 2$ and $\ell = 3$) has the highest probability of matching the data. These estimates also agree with the time at which the frequency and damping time of the fundamental quadrupolar mode match values obtained from the data analysis of GW150914 \cite{ligo-testsGR,brito-2018oct}. 

\begin{figure*}[htb!]
	\centering
	\includegraphics[width=0.75\linewidth]{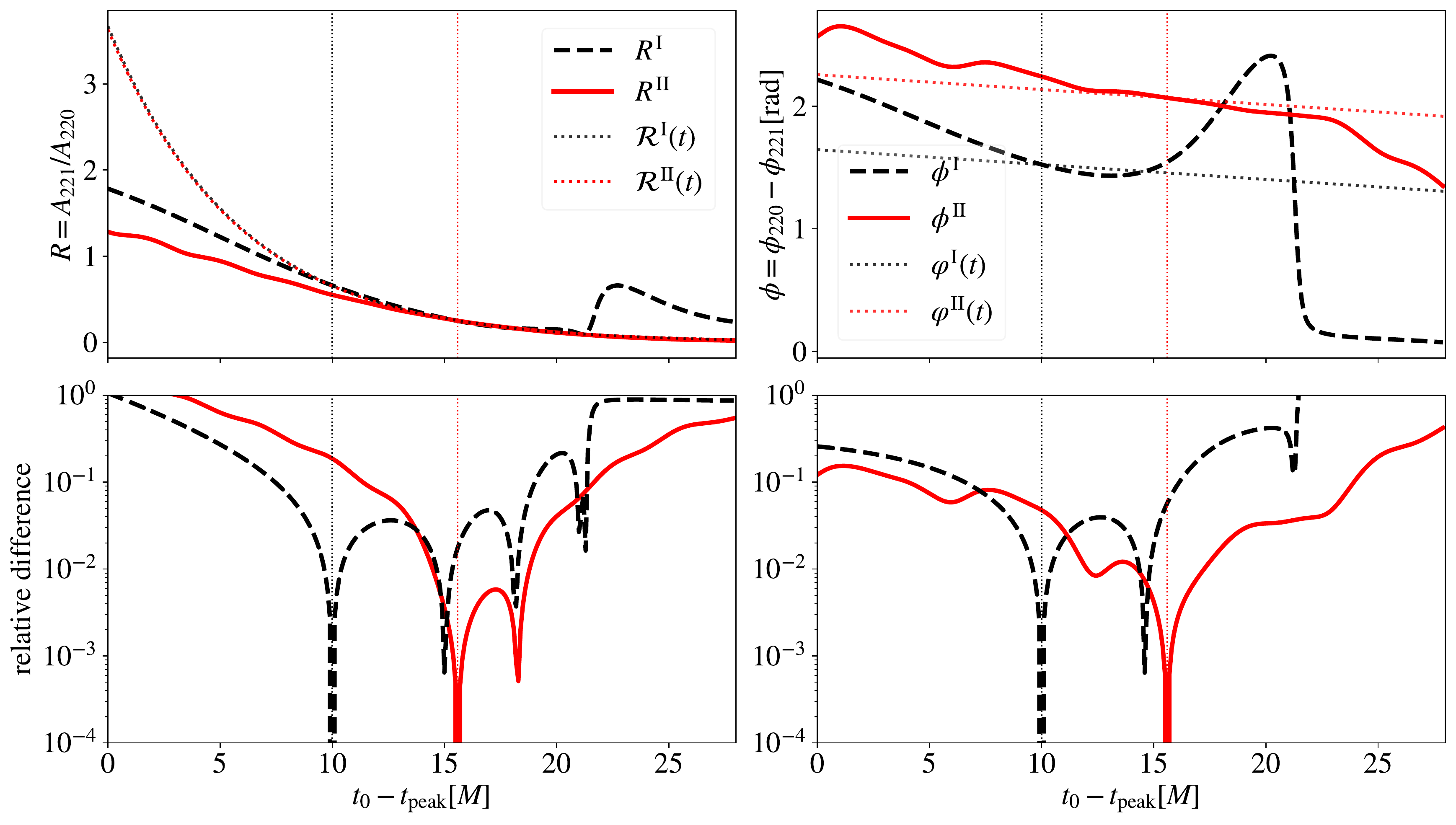}
	\caption{\textit{Top left:} Amplitude ratio $R^{\alpha} = A_{221}/A_{220}$ between the first overtone and the fundamental mode as a function of the initial time, where $\alpha$ labels methods I and II as defined in Figure \ref{fig:mismatch} and in the text. The dotted curves show the expected time-dependence in the linear regime $\mathcal{R}^{\alpha}(t)$ (\ref{eq:R(t)}) adjusted to the best initial times $t_0^{\alpha}$ as obtained in Figure \ref{fig:mismatch}, indicated with vertical lines.
	\textit{Bottom left:} Relative difference between the fitted ratio $R^\alpha$ and the expected time-dependent ratio $\mathcal{R}^\alpha$. The increasing difference between $R^{\alpha}$ and $\mathcal{R}^\alpha$ towards $t_{\rm peak}$ may be due to non-negligible contributions of higher overtones or non-linear dynamics in the waveform, while the increasing errors for larger $t$ come from the vanishing of $R^{\alpha}$ at later times. 
	\textit{Top and bottom right:} Same as in the left panels, but for the phase difference $\phi^\alpha = \phi_{220} - \phi_{221}$ between the first overtone and the fundamental mode and the expected time dependence in the linear regime $\varphi^\alpha(t)$ (\ref{eq:phi(t)}).}
	\label{fig:R_avg}
\end{figure*}

Figure \ref{fig:R_avg} shows in the two upper plots the values obtained with our two different fits for the initial amplitude ratio $R$ (left) and initial phase difference $\phi$ (right) as a function of the initial time $t_0$. The dotted vertical lines show again the best initial times obtained from methods I and II, see Figure \ref{fig:mismatch}. 
Both $\phi$ and $R$ decrease with the initial time $t_0$, but we note that method I is more sensitive and presents unphysical variations after the best initial time as $R$ approaches zero (and consequently $\ddot{\theta}_{22} \to 0$). A similar behavior is also present for $\phi$ obtained with method II at slightly later times, not shown in the plot. 

Even if our estimates for the initial time $t_0$ do not coincide with the initial time of the ringdown, we expect that the amplitude ratio calculated at $t_0^\alpha$ (where $\alpha = {\rm I, II}$ labels the method used) should be correct. As long as the linear regime is valid, the amplitude ratio as a function of time can be written as 
\begin{equation}
\mathcal{R}^\alpha(t) = R^\alpha(t_0^{\alpha})e^{(\omega_{221}^i - \omega_{220}^i) (t_0^{\alpha} - (t-t_\mathrm{peak}))},
\label{eq:R(t)}
\end{equation}
where $R^\alpha(t_0^\alpha)$ is the fitted amplitude ratio $R^\alpha$ at the best initial time $t_0^\alpha$ for each fit. Similarly, we can also write the phase difference as a function of time
\begin{equation}
\varphi^\alpha (t) = \phi^\alpha(t_0^\alpha) - (\omega^r_{220} - \omega^r_{221})\left((t - t_\mathrm{peak}) - t_0^\alpha\right).
\label{eq:phi(t)}
\end{equation}
Expressions (\ref{eq:R(t)}) and (\ref{eq:phi(t)}) are also represented in the upper plots of Figure \ref{fig:R_avg} as dotted curves. Our results show very good agreement for $R$ between the two methods around the best initial times $t_0^{\alpha}$, with larger deviations observed for $\phi$. 
In the lower plots, the relative difference between the fits and the expressions for $\mathcal{R}^\alpha$ and $\varphi^\alpha$ is shown as a function of time. In both plots the larger differences at earlier times indicate that near $t_{\rm peak}$ the non-linear dynamics or the higher overtones have significant contributions in the waveform, while the larger differences at later times are caused by the exponential vanishing of $R$.

\begin{figure*}[htb!]
	\centering
	\includegraphics[width=0.75\linewidth]{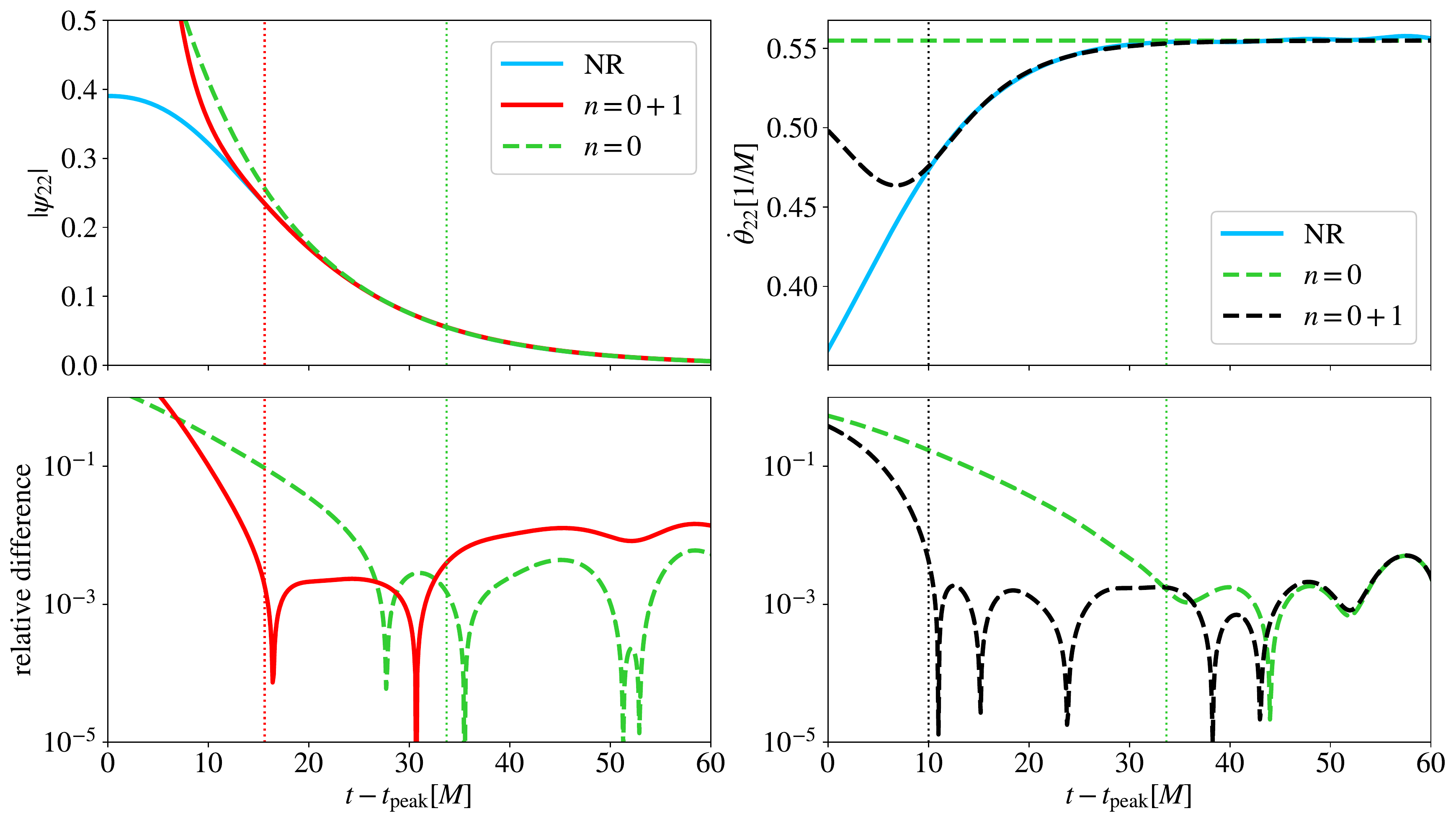}
	\caption{{\it Top:} Amplitude of the ringdown waveform $|\psi_{22}|$ (left) and time derivative of the phase $\dot{\theta}_{22}$ (right) as a function of time for the SXS:BBH:0305 NR simulation, together with our best fits for the ringdown considering only the fundamental mode ($n = 0$) and the fundamental mode + first overtone ($n = 0 + 1$), using the best initial times $t_0^{\alpha}$ found with methods I (right) and II (left). {\it Bottom:} Relative differences between the fits and the simulation as a function of time. The dotted vertical lines indicates the initial times for each fit, as shown before in Figures \ref{arctan_peak0} and \ref{fig:mismatch}.}
	\label{fig:fits}
\end{figure*}

Finally, in Figure \ref{fig:fits} we present a comparison between the simulation SXS:BBH:0301 (solid blue) and our best fits for the fundamental mode and for the fundamental mode + first overtone. The fits are performed at the best initial times ($t_{n = 0}$, $t_0^{\rm I}$ and $t_0^{\rm II}$) shown with vertical lines. Both for the waveform $\psi_{22}$ and for the phase derivative $\dot{\theta}_{22}$ we can see that adding the first overtone pushes back the best initial time for the fit, but the residuals are comparable after $t_{n=0}$. 

In summary, we have found out that the waveform has a non negligible contribution from the first overtone, with an amplitude ratio $R = A_{221}/A_{220} = 0.66$ at $t_0^{\rm I} = t_\mathrm{peak} + 10M$. So far, most spectroscopic analyses have neglected the overtones and focused instead on higher harmonic modes. However, higher harmonic modes have very low excitation amplitudes relative to the quadrupolar mode. In the example we have considered so far, we have found that the amplitude ratios corresponding to the next higher harmonics are $A_{210}/A_{220} = 0.05$, $A_{330}/A_{220} = 0.07$, $A_{440}/A_{220} = 0.04$.

\subsection{Detectability and resolvability}
\label{sec:resolve}

The higher excitation amplitude of the $(2,2,1)$ mode seems to indicate that the first overtone of the quadrupolar mode is more promising than the higher harmonics to test the no-hair theorem for the representative case considered so far. However, we also have to take into account that the first overtone decays about three times faster than the fundamental higher harmonics, which damp at a rate similar to the fundamental quadrupolar mode. In this subsection we present a preliminary assessment of these points.

\begin{figure}[htb!]
	\centering
	\includegraphics[scale = 0.31]{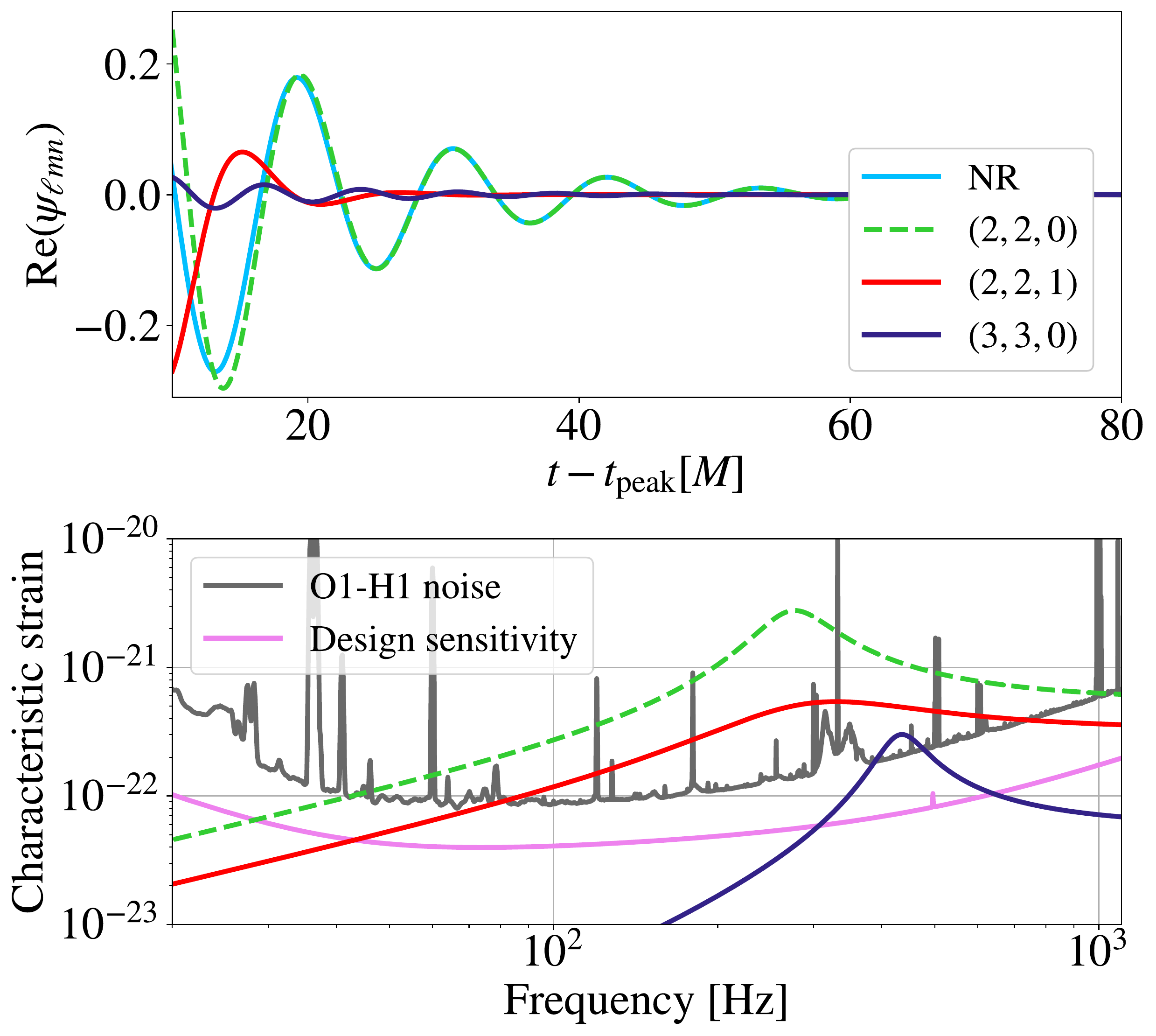}
	\caption{{\it Top:} The real part of the NR waveform and the fitted waveforms of the modes $(2,2,0)$, $(2,2,1)$ and $(3,3,0)$. We used the parameters obtained with method II to plot the $(2,2,1)$ mode. At early times, the first overtone of the quadrupolar mode $(2,2,1)$ has a higher amplitude than the most dominant higher harmonic mode $(3,3,0)$. {\it Bottom:} Characteristic strain (defined as in eq. (17) of \cite{Moore:2014lga}) of the modes rescaled with the remnant mass and luminosity distant of GW150914, $M_f =  63 M_\odot$ and $D_L = 440 \text{Mpc}$, and the characteristic noise curve of the Advanced LIGO Hanford detector during O1 (gray) and at design sensitivity (violet).}
	\label{fig:3-modes}
\end{figure}

In the upper plot of Figure \ref{fig:3-modes} we present the real part of the waveforms of the $(2,2,0)$, $(2,2,1)$ and $(3,3,0)$ modes, calculated using method II. At early times the amplitude of the $(2,2,1)$ mode is considerably higher than the amplitude of the $(3,3,0)$, which is the most dominant higher harmonic mode in this case as we have showed above.

Rescaling our best fits for the waveform with the remnant mass and luminosity distant of GW150914 \cite{LIGOScientific:2018mvr}, $M_f =  63 M_\odot$ and $D_L = 440 \text{Mpc}$, we can calculate the SNR $\rho_{\ell mn}$ of the plus polarization (real part) of the $(2,2,0)$, $(2,2,1)$ and $(3,3,0)$ modes using the Advanced LIGO Hanford detector noise curve during O1 \cite{GW150914-data}, as indicated in the lower plot of Figure \ref{fig:3-modes}. We obtain $\rho_{220} \sim 12.1$, $\rho_{221} \sim 3.2$ and $\rho_{330} \sim 0.7$. Therefore we find $\rho_{221} \sim 4.5 \rho_{330}$, even though the (2,2,1) mode damps about three times faster than the (3,3,0) mode.

A QNM is detectable in the ringdown if it has SNR greater than some threshold, say $\rho_{\ell mn} \gtrsim 5$. Nevertheless, the SNR calculation is not enough to determine whether the mode frequencies can be distinguished in the detection, in particular because the oscillation frequencies of the (2,2,0) and (2,2,1) modes are very close (see Table \ref{tabelafreqn}). According to the Rayleigh criterion  proposed by \cite{Berti-2006mar}, the frequencies of oscillation $f_{\ell m n} = \omega_{\ell m n}^r/(2\pi)$ and damping times $\tau_{\ell m n} = 1/\omega_{\ell m n}^i$ of two modes with $(\ell,m,n) \ne (\ell',m',n')$ are resolvable if they satisfy
\begin{align}
	\Delta f_{\ell m n,\ell'm'n'} \equiv & 
	\abs{f_{\ell m n} - f_{\ell^\prime m^\prime n^\prime}}> \max(\sigma_{f_{\ell m n}}, \sigma_{f_{\ell^\prime m^\prime n^\prime}}),\nonumber \\
	\Delta \tau_{\ell m n,\ell'm'n'} \equiv & 
	\abs{\tau_{\ell m n} - \tau_{\ell^\prime m^\prime n^\prime}} > \max(\sigma_{\tau_{\ell m n}}, \sigma_{\tau_{\ell^\prime m^\prime n^\prime}}),
	\label{eq:rayleight}
\end{align}
where the errors $\sigma_f$ and $\sigma_{\tau}$ in the frequencies and damping times can be estimated numerically using the Fisher matrix formalism \cite{Finn:1992wt}.

For our analysis of an event similar to GW150914, the Rayleigh criterion indicates that the frequencies of the (2,2,0) and (2,2,1) modes are not resolvable, but their damping times are; the minimum ringdown SNR needed for resolving the damping times of these two modes is $\rho_{220,221} \sim 1.5$. On the other hand, we find that the damping times of the $(2,2,0)$ and $(3,3,0)$ modes are not resolvable, but their frequencies are; the minimum SNR needed for resolving the frequencies is $\rho_{220,330} \sim 5.6$.

We also estimate the minimum ringdown SNR required to resolve \textit{both}  the frequencies and damping times. We find that the SNR needed to resolve the $(2,2,0)$ and $(2,2,1)$ modes is $\rho_{220,221} \gtrsim 55$, whereas for the $(2,2,0)$ and $(3,3,0)$ modes we must have $\rho_{220,330} \gtrsim 780$ when using the Advanced LIGO design sensitivity noise curve \cite{Design-sensitivity}. (We note that our result for the overtone resolvability is compatible with \cite{Bhagwat:2019dtm} if we start our analysis at $t_0 = t_\mathrm{peak}$ instead of $ t_\mathrm{peak} + 10M$.)

It is important to stress that this analysis relies on a Fisher matrix estimation of the statistical errors, which are expected to be larger for the actual parameter estimation of observed data. However, our results indicate that the first overtone could have an excitation amplitude high enough to be seen in data analyses of the ringdown phase of the strongest events, as indicated by the preliminary work done in \cite{Isi:2019aib}, whereas detections of higher harmonic modes in the LIGO/Virgo data seem to be less favored \cite{carullo-2019jun}.

\section{Mass ratio dependence}
\label{sec:mass_ratio}

Now we will systematically explore SXS simulations with increasing mass ratio $q \equiv M_1/M_2$ and initial zero-spin, in order to assess how relevant a contribution the first overtone will have in binary black hole systems with non-equal mass ratios.  

\begin{figure*}[htb!]
	\centering
	\includegraphics[width=0.75\linewidth]{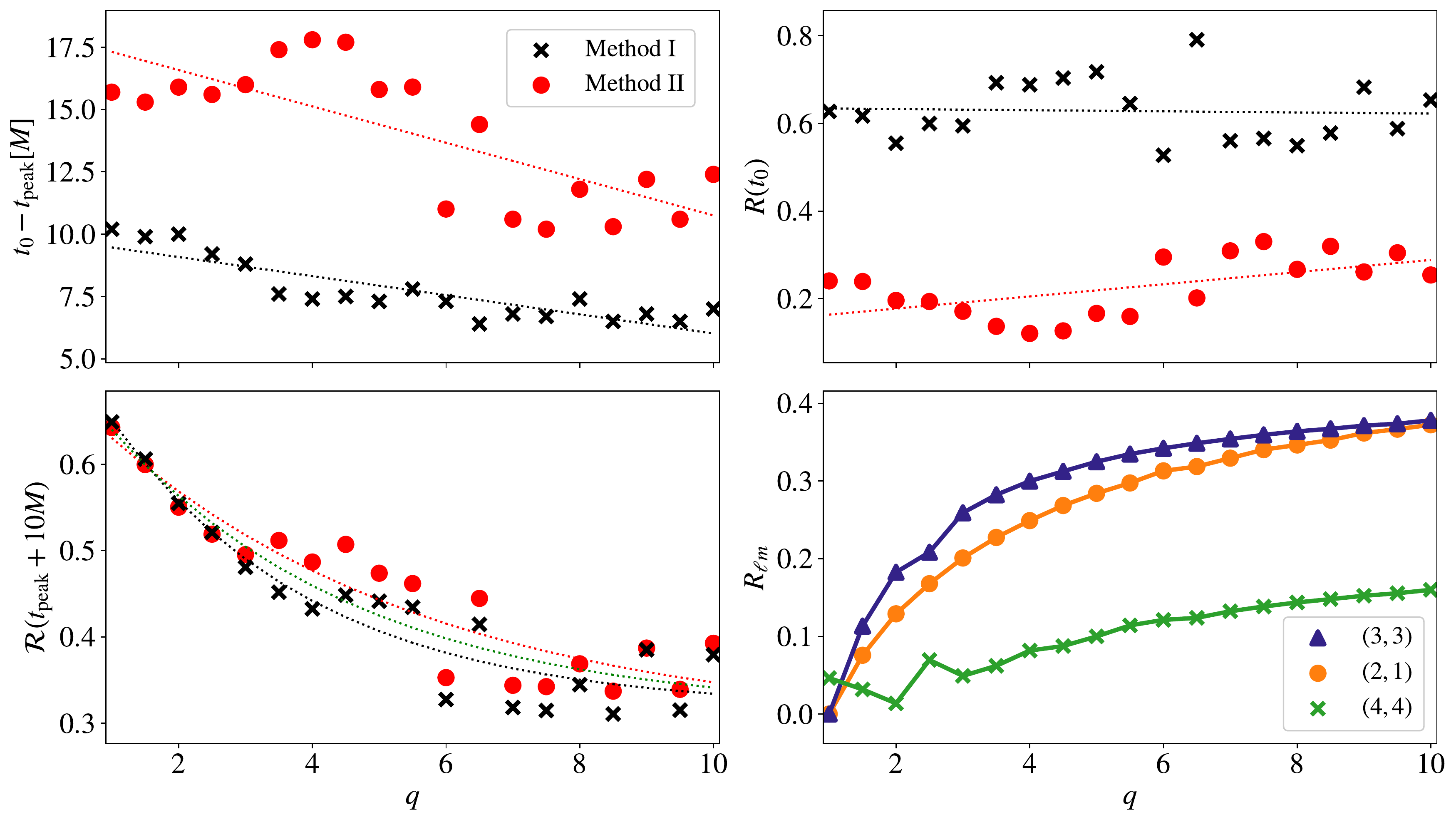}
	\caption{{\it Top:} Initial times $t_0$ (left) and amplitude ratio $R = A_{221}/A_{220}$ determined at $t_0$ (right) as a function of the binary mass ratio $q$, obtained with methods I and II (defined in Figure \ref{fig:mismatch}). The ringdown waveform is well described by the fundamental mode and the first overtone at earlier times for higher mass ratios. {\it Bottom:} Amplitude ratio $\mathcal{R}(t)$ as defined before (left) and amplitude ratio $R_{\ell m} = A_{\ell m 0}/A_{220}$ between the fundamental mode of the higher harmonic modes $(\ell, m) = (2, 1)$, $(3, 3)$ and $(4, 4)$ and the fundamental quadrupolar mode $(2, 2)$ (right), both evaluated at $t = t_{\mathrm{peak}} + 10M$, as a function of the mass ratio. Methods I and II show good agreement, once corrected to the same fiducial time, and the green dotted curve shows the best combined fit for an exponential decay function. $R_{\ell m}$ is approximately independent of the initial time. The first overtone has a higher amplitude than all harmonic mode for lower mass ratios $(q \lesssim 5)$ and is comparable to the modes $(2,1)$ and $(3,3)$ for higher mass ratios.}
	\label{fig:R_n_lm}
\end{figure*}

We have reproduced our analysis for a set of 19 simulations with $q$ ranging from 1 to 10. The upper left  plot in Figure \ref{fig:R_n_lm} shows the initial time $t_0$ as a function of the binary mass ratio $q$. We can see that the waveform is well described by the fundamental mode and the first overtone at earlier times for higher mass ratios: as the binary mass ratio increases, the linear perturbation regime is approached closer to the merger and the contributions of the overtones become less relevant. 

In the upper right plot of Figure \ref{fig:R_n_lm} we present the amplitude ratio $R$ at $t_0$ as a function of the binary mass ratio $q$. The observed spread between the results of methods I and II is mostly explained by the dependence of $t_0$ on $q$, and the results obtained with both methods can be nicely unified when we present the expected values obtained with  eq.~(\ref{eq:R(t)}) at the same fiducial time $t = t_{\mathrm{peak}} + 10M$ in the lower left plot. The green dotted curve is the best fit for an exponential decay function, taking into account the results from both methods I and II:
\begin{equation}
\mathcal{R}(q, t_\mathrm{peak} + 10M) = 0.4 e^{-0.3q} + 0.3, 
\end{equation}
where the asymptotic amplitude ratio for large $q$ ($\mathcal{R} \to 0.3$) is compatible with the point particle limit at $t = t_\text{peak} + 10 M$. (We thank Vitor Cardoso for pointing this out.)

The lower right plot of Figure \ref{fig:R_n_lm} shows the amplitude ratio $R_{\ell m}$ between the fundamental mode of the higher harmonic modes $(\ell, m) = (2, 1)$, $(3, 3)$ and $(4, 4)$ and the fundamental quadrupolar mode $(2, 2)$ at $t = t_{\mathrm{peak}} + 10M$. We can see that the first overtone (2,2,1) has a higher amplitude ratio $\mathcal{R}$ than all harmonic modes for lower mass ratios $q \lesssim 5$. For higher mass ratios, the asymptotic value of $\mathcal{R}$ is comparable to the $(2,1,0)$ and $(3,3,0)$ values. 
We also note that $R_{\ell m}$ does not depend on the initial time due to similar damping times between the fundamental modes with different $(\ell,m)$ (see Table \ref{table-lm}), and that our results for the higher harmonics are in good agreement with those of Fig. 1 of \cite{Cotesta-2018oct}. 

\begin{table}[htb!]
\centering
\caption{Ringdown frequencies of the fundamental mode for the first harmonics of a remnant black hole with mass $M_f = 0.9540 M$ and final spin $a = 0.6556 M$, resulting from the merger of an equal mass $(q = 1)$ non-spinning binary.}
\begin{ruledtabular}
\begin{tabular}{c c c }
$(\ell, m, n)$ & $M\omega^r_{\ell m n}$ & $M\omega^i_{\ell m n}$\\ \hline
$(2,2,0)$ & $0.5524$ & $0.0852$ \\
$(2,1,0)$ & $0.4742$ & $0.0865$ \\
$(3,3,0)$ & $0.8757$ & $0.0874$ \\
$(4,4,0)$ & $1.1861$ & $0.0889$
\end{tabular}
\end{ruledtabular}
\label{table-lm}
\end{table}

\begin{figure}[htb!]
	\centering
	\includegraphics[width=0.95\linewidth]{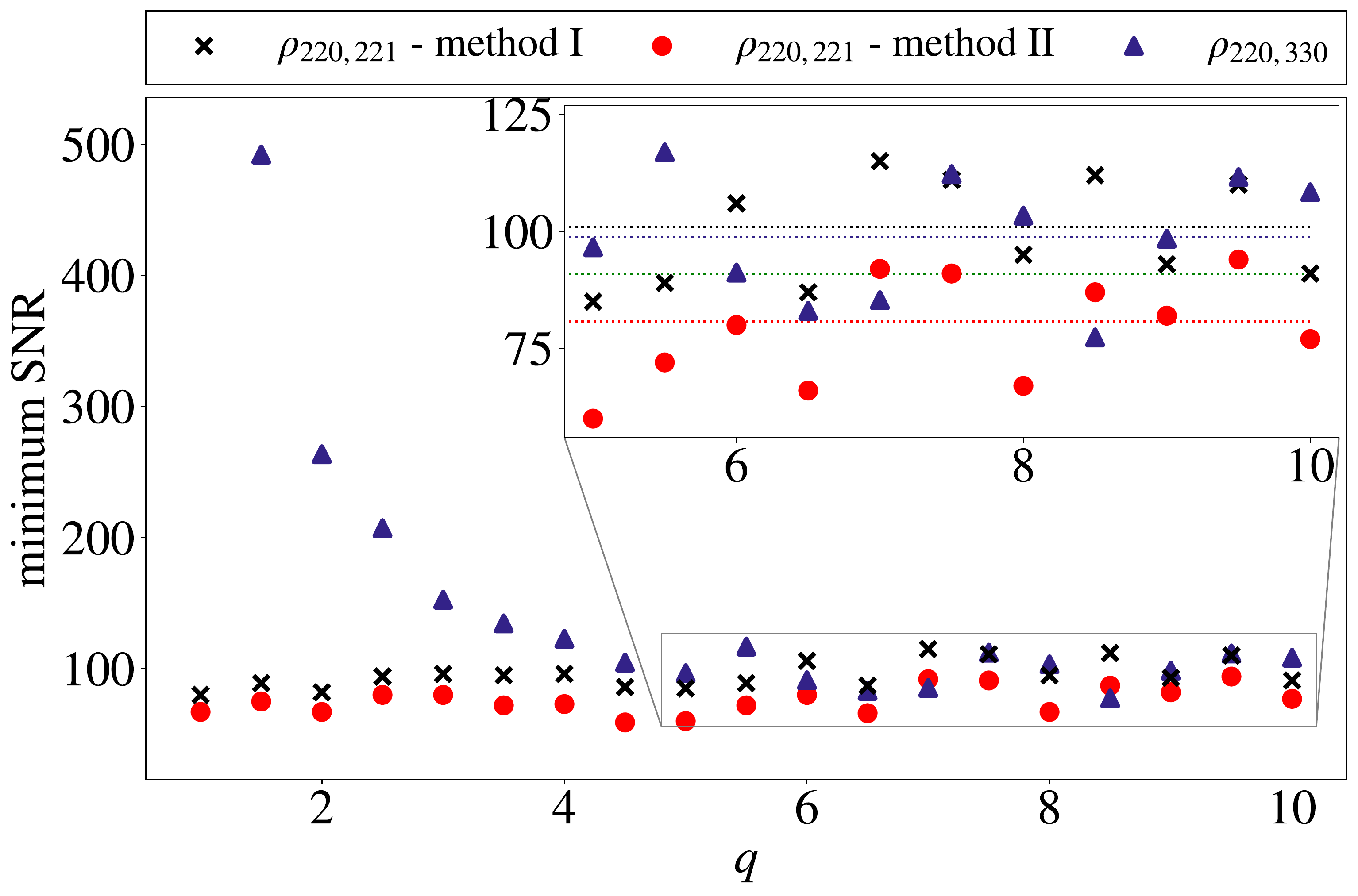}
	\caption{Minimum ringdown SNR required to resolve both the frequencies and damping times of the fundamental quadrupolar mode (2,2,0) mode and the first overtone (2,2,1) (red dots and black crosses) and the fundamental higher harmonic mode (3,3,0) (blue triangles) as function of the binary mass ratio $q$, keeping the mass of the black hole merger remnant consistent with GW150914. The green dotted line shows the average $\rho_{220,221}$ using the results from both methods I and II. The minimum SNR needed to resolve the (2,2,1) mode is only weakly dependent on $q$ (see text) and it is consistently lower than the minimum SNR needed for resolving the (3,3,0). For $q \gtrsim 5$, $\rho_{220,221}$ is 8\% lower than $\rho_{220,330}$.}
	\label{fig:rho_crit}
\end{figure}

The preliminary Fisher Matrix analysis presented in subsection \ref{sec:resolve} can be extended to the case of larger mass ratios. The results reported in Figure \ref{fig:rho_crit} were obtained by keeping the final mass of the remnant black hole compatible with GW150914 and using the Advanced LIGO design sensitivity noise curve. This analysis takes into account the amplitude of the modes, reported in Figure \ref{fig:R_n_lm}, but also the difference between the frequencies and damping times of the two modes, see eq. (\ref{eq:rayleight}).

Non-spinning binaries with more unequal masses (larger $q$) produce a remnant black hole with lower spin. A lower black hole spin increases $\Delta f_{220,221}$ \cite{Berti-2006mar}, making the two modes easier to resolve. The increasing $\Delta f_{220,221}$ combined with the decreasing relative amplitude of the (2,2,1) mode results in the weak dependence of the minimum ringdown SNR $\rho_{220,221}$ needed for resolvability on $q$. In contrast, increasing $q$ leads to a higher relative amplitude of the (3,3,0) mode \emph{and} to a larger $\Delta \tau_{220,330}$ (also due to the lower remnant spin); both effects lead to the decrease in $\rho_{220,330}$ seen in Figure \ref{fig:rho_crit}.

Consequently we can conclude that even for highly unequal binaries with $q \gtrsim 5$ the overtone mode (2,2,1) should be more easily resolvable (and therefore detectable) than the higher harmonic (3,3,0), with an averaged minimum SNR for resolvability $\rho_{220,221} \sim 91$, approximately 8\% lower than the corresponding averaged $\rho_{220,330}$.

But it is important to note that the \emph{inclination} of the binary is also relevant for our discussion.
For the example of a wave coming from an optimal direction\footnote{Such that the detector pattern functions become $F_+ = 1$ and $F_{\times} = 0$ \cite{maggiore2008gravitational}.}, the full signal observed by a detector will be $\sum_{\ell m} Y_{\ell m}(\iota,\varphi)\Re[\psi_{\ell m}]$, with $Y_{\ell m}$ the -2 spin-weighted spherical harmonics and $\iota$ and $\varphi$ the inclination and azimuth angles of the binary, respectively \cite{Boyle:2019kee,berti.cardoso.casals}. 

The effective amplitude of each mode in the detector band will be roughly $|Y_{\ell m}(\iota,\varphi)A_{\ell mn}|$ and a reference value for  $|Y_{\ell m}(\iota,\varphi)|$ should be given by its average over all possible directions and inclinations. The ratio of angular averages $\langle|Y_{22}(\iota,\varphi)|\rangle/\langle|Y_{33}(\iota,\varphi)|\rangle$ is approximately $0.94$ which leaves our previous conclusions almost unchanged.

However, events viewed \emph{face-on} ($\iota = 0$) are stronger than events viewed \emph{edge-on} ($\iota = \pi/2$). Otherwise identical events with a given SNR will be detected out to a distance (see section 7.7.2 of \cite{maggiore2008gravitational})
$$d_{\rm sight} \propto \left|\frac{1+\cos^2\iota}{2}\right|,$$ 
which is 2 times larger if they are viewed face-on than if they are viewed edge-on. Therefore, the expected number of detections as a function of the inclination angle of the binary is given by
$$ 
N(\iota) = \left(\frac{\textrm{event rate}}{\rm volume}\right) \times \textrm{(observation time)} \times V(\iota),
$$
where $V(\iota) \propto d^3_{\rm sight}$ is the observable volume for a given SNR. With these considerations, we conclude that the expected amplitude ratio between the (2,2,1) and the (3,3,0) modes (as inferred form Figure \ref{fig:R_n_lm}) should be boosted by a factor 
$\langle|N(\iota)Y_{22}(\iota,\varphi)|\rangle/\langle|N(\iota)Y_{33}(\iota,\varphi)|\rangle \sim 1.5$. 

\begin{figure}[htb!]
	\centering
	\includegraphics[width=0.95\linewidth]{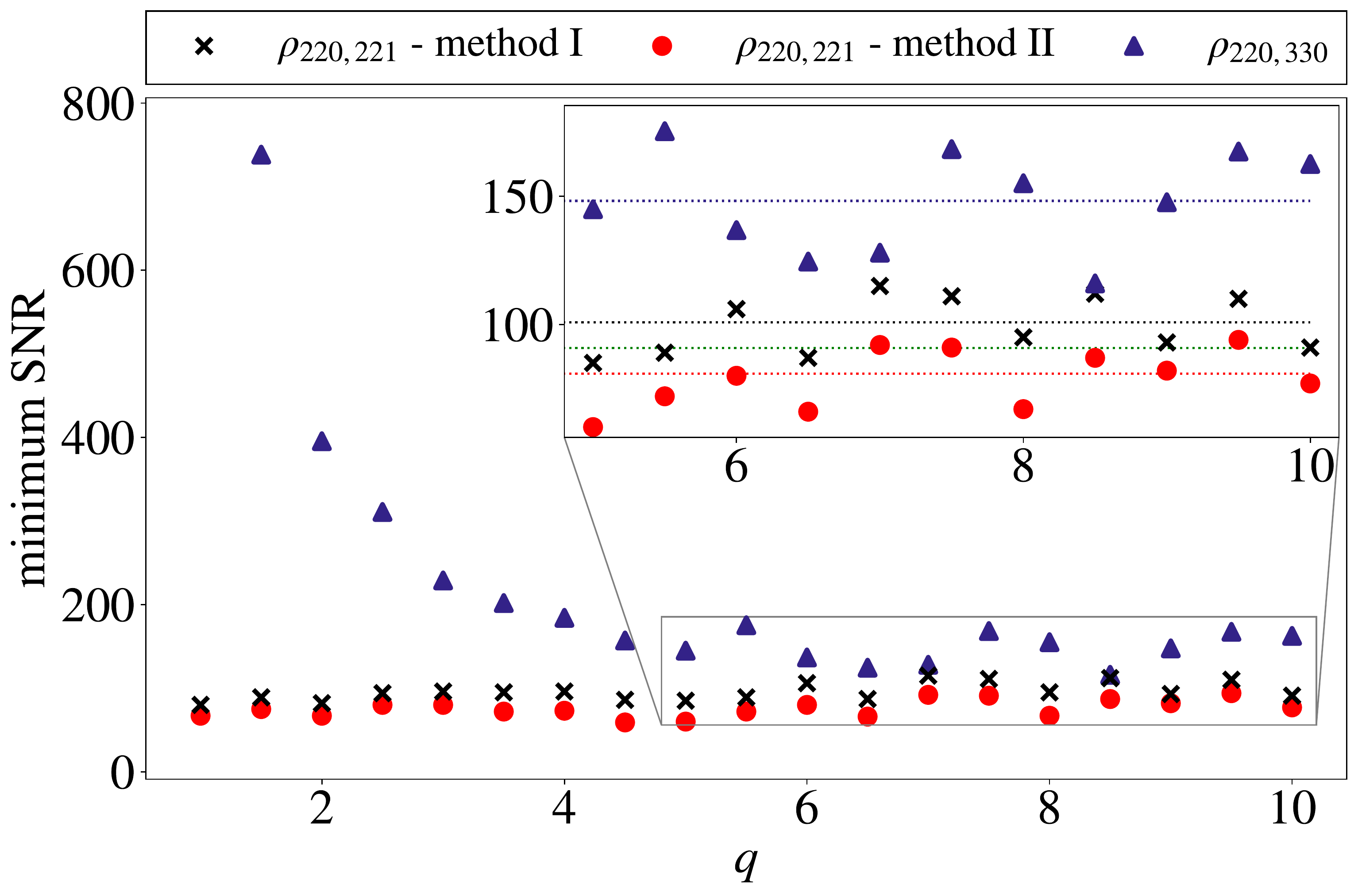}
	\caption{Same as Figure \ref{fig:rho_crit}, but corrected for the observed amplitude of the mode at the detector and weighted by the expected number of detections (see text). Both factors are functions of the inclination angle of the binary and are averaged over all possible inclinations. For $q \gtrsim 5$, $\rho_{220,221}$ is approximately 40\% lower than $\rho_{220,330}$.}
	\label{fig:rho_crit_rY}
\end{figure}

This correction quantifies the observational preference for events in which the (2,2,1) mode is more promising for detection than the (3,3,0) mode. It does not change our previous results for $\rho_{220,221}$, but it does increase the averaged $\rho_{220,330}$ by 50\%, as we can see in Figure \ref{fig:rho_crit_rY}. As a result, the averaged $\rho_{220,221}$ is approximately 40\% lower than the corresponding $\rho_{220,330}$ for $q \gtrsim 5$.  Therefore, we can conclude that the (2,2,1) mode is more promising for detection than the (3,3,0) mode, even for binaries with a mass ratio much larger than any GW events reporter so far.

\section{Conclusions}
\label{sec:conclusions}

We have used NR simulations from the Simulating eXtreme Spacetimes project (SXS) \cite{Boyle:2019kee,SXS-catalog} to estimate the contribution of overtones and higher harmonics in the ringdown of a BBH merger, with the aim to identify the most promising route for observationally testing the no-hair theorem using gravitational wave detections and black hole spectroscopy. 

Initially we focused on the quadrupolar mode and we used the waveform $\psi_{22}$ and the time derivative of the  phase $\theta_{22}$ in two different methods to determine the initial time $t_0$ from which the simulated data is well described by the $(2,2,0)$ and the $(2,2,1)$ modes. For the nearly-equal mass case we found a initial time in agreement with previous ringdown analyses \cite{Carullo-2018nov,carullo-2019jun,thrane-2017nov}. Additionally, we found that the initial times obtained decrease with increasing mass ratio of the binary system.

By scaling the excitation amplitudes to a fiducial time $t_{\rm peak}+10M$ we found  that the (2,2,1) mode will always be more significant than the fundamental higher harmonic modes for BBH systems with low mass ratios, from 1:1 to approximately 5:1. In particular, for an event similar to GW150914 we have $\mathcal{R}(t_\mathrm{peak} + 10M) = A_{221}/A_{220} = 0.66$, more than 10 times larger than the amplitudes of the other harmonics. For mass ratios larger than 5:1, our results indicate an interesting ``equipartition'' and the (2,2,1), (2,1,0) and (3,3,0) modes have comparable amplitudes of approximately $0.35 A_{220}$. The amplitude ratio of the first overtone and the fundamental mode of the quadrupolar harmonic seems to asymptote with increasing mass ratio to a constant value $\mathcal{R} = 0.3$, which is compatible with the point particle limit.

Almost all of the GW detections reported so far are compatible with equal-mass black hole binaries \cite{LIGOScientific:2018mvr,venumadhav2019new,zackay2019detecting}\footnote{ A recent analysis of GW170729 \cite{Chatziioannou:2019dsz} showed that the event is inconsistent with equal-mass binary at the 90\% level with a preferred mass ratio in the range 1.25 - 3.33.}. Therefore our results indicate a promising prospect for using the GW data to test the no-hair theorem with overtones, even though it is expected that in O3 and future observing runs some events with a higher mass ratio may also be detected. 

However, the close frequencies of the (2,2,0) and (2,2,1) modes and the faster damping time of the overtone will necessarily make this a challenging detection, as we showed in our preliminary analysis in subsection \ref{sec:resolve}, using a Fisher matrix analysis to implement the Rayleigh criterion. For a simulation consistent with GW150914, the minimum ringdown SNR needed to resolve both the frequencies of oscillation and damping times of the fundamental mode $(2,2,0)$ and its first overtone $(2,2,1)$ is $\rho_{220,221} \sim 55$, if the ringdown analysis starts at $t_{\rm peak} + 10M$. However, the damping times alone are different enough that they should be resolvable if both modes are detectable, indicating that this could be a promising feature to look for in future data. 

We have also extended the resolvability analysis to unequal mass ratios (always keeping the mass of the final black hole compatible with GW150914), taking  into account the dependence of the observed mode amplitudes at the detector on the binary inclination angle, weighted by the expected number of detections. We were able to conclude that the detection of the (2,2,1) mode is favored over the (3,3,0) mode for the entire range of mass ratios considered in this work. For binaries with final mass compatible with GW150914 and mass ratio larger than 5:1, the (2,2,1) mode has a minimum ringdown SNR for resolvability ${\rho_{220,221}} \sim 91$, approximately 40\% lower than the corresponding SNR for the (3,3,0) mode.

A coherent mode stacking analysis \cite{Yang_2017, Berti_2018} may be needed to improve the significance of the first overtone in a Bayesian model comparison, which we expect to perform with the already reported detections. We are also working on extending our analysis to cases with non-zero initial spins and eccentricities (we considered here only one case with nonzero spins in Sections \ref{sec:fundamental} and \ref{sec:overtones}), as it is well known that initial spin affects the higher harmonics excitation amplitudes \cite{Cotesta-2018oct}.

Initial evidence for the observation of the first overtone was recently reported by \cite{Isi:2019aib},  with frequency results consistent with the no-hair theorem at the $\sim 20 \%$ level.
It is likely that a more precise identification of the (2,2,1) mode in the GW data will have to wait for more signals with higher SNR. However, this is only a matter of time, and the development of the necessary analysis tools and theoretical understanding is timely. We need to be prepared for the surprises that will undoubtedly come from GW astronomy.

\begin{acknowledgments}
We thank Emanuele Berti, Vitor Cardoso, Gregorio Carullo, Xisco Forteza, Max Isi, Luis Lehner, Luciano Rezzolla and Erik Schnetter for useful discussions and comments on our work. We are especially thankful to Bernard Kelly for his help in the initial stages of this project and to Cole Miller for suggesting the weighting of the amplitude ratios by the expected number of detections. 
IO was partially supported by grant 2018/21286-3 of the S\~ao Paulo Research Foundation (FAPESP) and by the Federal University of ABC. CC acknowledges support from grant 303750/2017-0 of the Brazilian National Council for Scientific and Technological Development (CNPq) and by NASA under award number 80GSFC17M0002. We are grateful for the hospitality of Perimeter Institute where part of this work was carried out. Research at Perimeter Institute is supported in part by the Government of Canada through the Department of Innovation, Science and Economic Development Canada and by the Province of Ontario through the Ministry of Economic Development, Job Creation and Trade. This research was also supported in part by the Simons Foundation through the Simons Foundation Emmy Noether Fellows Program at Perimeter Institute.
\end{acknowledgments}

\bibliography{main.bib}

\begin{thebibliography}{50}%
\makeatletter
\providecommand \@ifxundefined [1]{%
 \@ifx{#1\undefined}
}%
\providecommand \@ifnum [1]{%
 \ifnum #1\expandafter \@firstoftwo
 \else \expandafter \@secondoftwo
 \fi
}%
\providecommand \@ifx [1]{%
 \ifx #1\expandafter \@firstoftwo
 \else \expandafter \@secondoftwo
 \fi
}%
\providecommand \natexlab [1]{#1}%
\providecommand \enquote  [1]{``#1''}%
\providecommand \bibnamefont  [1]{#1}%
\providecommand \bibfnamefont [1]{#1}%
\providecommand \citenamefont [1]{#1}%
\providecommand \href@noop [0]{\@secondoftwo}%
\providecommand \href [0]{\begingroup \@sanitize@url \@href}%
\providecommand \@href[1]{\@@startlink{#1}\@@href}%
\providecommand \@@href[1]{\endgroup#1\@@endlink}%
\providecommand \@sanitize@url [0]{\catcode `\\12\catcode `\$12\catcode
  `\&12\catcode `\#12\catcode `\^12\catcode `\_12\catcode `\%12\relax}%
\providecommand \@@startlink[1]{}%
\providecommand \@@endlink[0]{}%
\providecommand \url  [0]{\begingroup\@sanitize@url \@url }%
\providecommand \@url [1]{\endgroup\@href {#1}{\urlprefix }}%
\providecommand \urlprefix  [0]{URL }%
\providecommand \Eprint [0]{\href }%
\providecommand \doibase [0]{http://dx.doi.org/}%
\providecommand \selectlanguage [0]{\@gobble}%
\providecommand \bibinfo  [0]{\@secondoftwo}%
\providecommand \bibfield  [0]{\@secondoftwo}%
\providecommand \translation [1]{[#1]}%
\providecommand \BibitemOpen [0]{}%
\providecommand \bibitemStop [0]{}%
\providecommand \bibitemNoStop [0]{.\EOS\space}%
\providecommand \EOS [0]{\spacefactor3000\relax}%
\providecommand \BibitemShut  [1]{\csname bibitem#1\endcsname}%
\let\auto@bib@innerbib\@empty
%</preamble>
\bibitem [{\citenamefont {Abbott}\ \emph {et~al.}(2019)\citenamefont {Abbott}
  \emph {et~al.}}]{LIGOScientific:2018mvr}%
  \BibitemOpen
  \bibfield  {author} {\bibinfo {author} {\bibfnamefont {B.~P.}\ \bibnamefont
  {Abbott}} \emph {et~al.} (\bibinfo {collaboration} {LIGO Scientific,
  Virgo}),\ }\href {\doibase 10.1103/PhysRevX.9.031040} {\bibfield  {journal}
  {\bibinfo  {journal} {Phys. Rev.}\ }\textbf {\bibinfo {volume} {X9}},\
  \bibinfo {pages} {031040} (\bibinfo {year} {2019})},\ \Eprint
  {http://arxiv.org/abs/1811.12907} {arXiv:1811.12907 [astro-ph.HE]}
  \BibitemShut {NoStop}%
%%CITATION = ARXIV:1811.12907;%%
\bibitem [{Gra()}]{GraceDB}%
  \BibitemOpen
  \href@noop {} {}\bibinfo {howpublished}
  {\url{https://gracedb.ligo.org/}}\BibitemShut {NoStop}%
\bibitem [{\citenamefont {Yunes}\ and\ \citenamefont
  {Siemens}(2013)}]{Yunes:2013dva}%
  \BibitemOpen
  \bibfield  {author} {\bibinfo {author} {\bibfnamefont {N.}~\bibnamefont
  {Yunes}}\ and\ \bibinfo {author} {\bibfnamefont {X.}~\bibnamefont
  {Siemens}},\ }\href {\doibase 10.12942/lrr-2013-9} {\bibfield  {journal}
  {\bibinfo  {journal} {Living Rev. Rel.}\ }\textbf {\bibinfo {volume} {16}},\
  \bibinfo {pages} {9} (\bibinfo {year} {2013})},\ \Eprint
  {http://arxiv.org/abs/1304.3473} {arXiv:1304.3473 [gr-qc]} \BibitemShut
  {NoStop}%
%%CITATION = ARXIV:1304.3473;%%
\bibitem [{\citenamefont {Abbott}\ \emph
  {et~al.}(2016{\natexlab{a}})\citenamefont {Abbott} \emph
  {et~al.}}]{ligo-testsGR}%
  \BibitemOpen
  \bibfield  {author} {\bibinfo {author} {\bibfnamefont {B.~P.}\ \bibnamefont
  {Abbott}} \emph {et~al.} (\bibinfo {collaboration} {LIGO Scientific,
  Virgo}),\ }\href {\doibase 10.1103/PhysRevLett.116.221101,
  10.1103/PhysRevLett.121.129902} {\bibfield  {journal} {\bibinfo  {journal}
  {Phys. Rev. Lett.}\ }\textbf {\bibinfo {volume} {116}},\ \bibinfo {pages}
  {221101} (\bibinfo {year} {2016}{\natexlab{a}})},\ \bibinfo {note} {[Erratum:
  Phys. Rev. Lett.121,no.12,129902(2018)]},\ \Eprint
  {http://arxiv.org/abs/1602.03841} {arXiv:1602.03841 [gr-qc]} \BibitemShut
  {NoStop}%
%%CITATION = ARXIV:1602.03841;%%
\bibitem [{\citenamefont {Barack}\ \emph {et~al.}(2019)\citenamefont {Barack}
  \emph {et~al.}}]{Barack_2019}%
  \BibitemOpen
  \bibfield  {author} {\bibinfo {author} {\bibfnamefont {L.}~\bibnamefont
  {Barack}} \emph {et~al.},\ }\href {\doibase 10.1088/1361-6382/ab0587}
  {\bibfield  {journal} {\bibinfo  {journal} {Classical and Quantum Gravity}\
  }\textbf {\bibinfo {volume} {36}},\ \bibinfo {pages} {143001} (\bibinfo
  {year} {2019})}\BibitemShut {NoStop}%
\bibitem [{\citenamefont {Kokkotas}\ and\ \citenamefont
  {Schmidt}(1999)}]{Kokkotas:1999bd}%
  \BibitemOpen
  \bibfield  {author} {\bibinfo {author} {\bibfnamefont {K.~D.}\ \bibnamefont
  {Kokkotas}}\ and\ \bibinfo {author} {\bibfnamefont {B.~G.}\ \bibnamefont
  {Schmidt}},\ }\href {\doibase 10.12942/lrr-1999-2} {\bibfield  {journal}
  {\bibinfo  {journal} {Living Rev. Rel.}\ }\textbf {\bibinfo {volume} {2}},\
  \bibinfo {pages} {2} (\bibinfo {year} {1999})},\ \Eprint
  {http://arxiv.org/abs/gr-qc/9909058} {arXiv:gr-qc/9909058 [gr-qc]}
  \BibitemShut {NoStop}%
%%CITATION = GR-QC/9909058;%%
\bibitem [{\citenamefont {Dreyer}\ \emph {et~al.}(2004)\citenamefont {Dreyer},
  \citenamefont {Kelly}, \citenamefont {Krishnan}, \citenamefont {Finn},
  \citenamefont {Garrison},\ and\ \citenamefont
  {Lopez-Aleman}}]{Dreyer-2004jan}%
  \BibitemOpen
  \bibfield  {author} {\bibinfo {author} {\bibfnamefont {O.}~\bibnamefont
  {Dreyer}}, \bibinfo {author} {\bibfnamefont {B.~J.}\ \bibnamefont {Kelly}},
  \bibinfo {author} {\bibfnamefont {B.}~\bibnamefont {Krishnan}}, \bibinfo
  {author} {\bibfnamefont {L.~S.}\ \bibnamefont {Finn}}, \bibinfo {author}
  {\bibfnamefont {D.}~\bibnamefont {Garrison}}, \ and\ \bibinfo {author}
  {\bibfnamefont {R.}~\bibnamefont {Lopez-Aleman}},\ }\href {\doibase
  10.1088/0264-9381/21/4/003} {\bibfield  {journal} {\bibinfo  {journal}
  {Class. Quant. Grav.}\ }\textbf {\bibinfo {volume} {21}},\ \bibinfo {pages}
  {787} (\bibinfo {year} {2004})},\ \Eprint
  {http://arxiv.org/abs/gr-qc/0309007} {arXiv:gr-qc/0309007 [gr-qc]}
  \BibitemShut {NoStop}%
%%CITATION = GR-QC/0309007;%%
\bibitem [{\citenamefont {Berti}\ \emph {et~al.}(2006)\citenamefont {Berti},
  \citenamefont {Cardoso},\ and\ \citenamefont {Will}}]{Berti-2006mar}%
  \BibitemOpen
  \bibfield  {author} {\bibinfo {author} {\bibfnamefont {E.}~\bibnamefont
  {Berti}}, \bibinfo {author} {\bibfnamefont {V.}~\bibnamefont {Cardoso}}, \
  and\ \bibinfo {author} {\bibfnamefont {C.~M.}\ \bibnamefont {Will}},\ }\href
  {\doibase 10.1103/PhysRevD.73.064030} {\bibfield  {journal} {\bibinfo
  {journal} {Phys. Rev.}\ }\textbf {\bibinfo {volume} {D73}},\ \bibinfo {pages}
  {064030} (\bibinfo {year} {2006})},\ \Eprint
  {http://arxiv.org/abs/gr-qc/0512160} {arXiv:gr-qc/0512160 [gr-qc]}
  \BibitemShut {NoStop}%
%%CITATION = GR-QC/0512160;%%
\bibitem [{\citenamefont {Brito}\ \emph {et~al.}(2018)\citenamefont {Brito},
  \citenamefont {Buonanno},\ and\ \citenamefont {Raymond}}]{brito-2018oct}%
  \BibitemOpen
  \bibfield  {author} {\bibinfo {author} {\bibfnamefont {R.}~\bibnamefont
  {Brito}}, \bibinfo {author} {\bibfnamefont {A.}~\bibnamefont {Buonanno}}, \
  and\ \bibinfo {author} {\bibfnamefont {V.}~\bibnamefont {Raymond}},\ }\href
  {\doibase 10.1103/PhysRevD.98.084038} {\bibfield  {journal} {\bibinfo
  {journal} {Phys. Rev.}\ }\textbf {\bibinfo {volume} {D98}},\ \bibinfo {pages}
  {084038} (\bibinfo {year} {2018})},\ \Eprint
  {http://arxiv.org/abs/1805.00293} {arXiv:1805.00293 [gr-qc]} \BibitemShut
  {NoStop}%
%%CITATION = ARXIV:1805.00293;%%
\bibitem [{\citenamefont {Baibhav}\ \emph {et~al.}(2018)\citenamefont
  {Baibhav}, \citenamefont {Berti}, \citenamefont {Cardoso},\ and\
  \citenamefont {Khanna}}]{Baibhav-2018feb}%
  \BibitemOpen
  \bibfield  {author} {\bibinfo {author} {\bibfnamefont {V.}~\bibnamefont
  {Baibhav}}, \bibinfo {author} {\bibfnamefont {E.}~\bibnamefont {Berti}},
  \bibinfo {author} {\bibfnamefont {V.}~\bibnamefont {Cardoso}}, \ and\
  \bibinfo {author} {\bibfnamefont {G.}~\bibnamefont {Khanna}},\ }\href
  {\doibase 10.1103/PhysRevD.97.044048} {\bibfield  {journal} {\bibinfo
  {journal} {Phys. Rev.}\ }\textbf {\bibinfo {volume} {D97}},\ \bibinfo {pages}
  {044048} (\bibinfo {year} {2018})},\ \Eprint
  {http://arxiv.org/abs/1710.02156} {arXiv:1710.02156 [gr-qc]} \BibitemShut
  {NoStop}%
%%CITATION = ARXIV:1710.02156;%%
\bibitem [{\citenamefont {Cardoso}\ \emph {et~al.}(2019)\citenamefont
  {Cardoso}, \citenamefont {Kimura}, \citenamefont {Maselli}, \citenamefont
  {Berti}, \citenamefont {Macedo},\ and\ \citenamefont
  {McManus}}]{Cardoso_2019}%
  \BibitemOpen
  \bibfield  {author} {\bibinfo {author} {\bibfnamefont {V.}~\bibnamefont
  {Cardoso}}, \bibinfo {author} {\bibfnamefont {M.}~\bibnamefont {Kimura}},
  \bibinfo {author} {\bibfnamefont {A.}~\bibnamefont {Maselli}}, \bibinfo
  {author} {\bibfnamefont {E.}~\bibnamefont {Berti}}, \bibinfo {author}
  {\bibfnamefont {C.~F.}\ \bibnamefont {Macedo}}, \ and\ \bibinfo {author}
  {\bibfnamefont {R.}~\bibnamefont {McManus}},\ }\href {\doibase
  10.1103/physrevd.99.104077} {\bibfield  {journal} {\bibinfo  {journal}
  {Physical Review D}\ }\textbf {\bibinfo {volume} {99}} (\bibinfo {year}
  {2019}),\ 10.1103/physrevd.99.104077}\BibitemShut {NoStop}%
\bibitem [{\citenamefont {McManus}\ \emph {et~al.}(2019)\citenamefont
  {McManus}, \citenamefont {Berti}, \citenamefont {Macedo}, \citenamefont
  {Kimura}, \citenamefont {Maselli},\ and\ \citenamefont
  {Cardoso}}]{McManus:2019ulj}%
  \BibitemOpen
  \bibfield  {author} {\bibinfo {author} {\bibfnamefont {R.}~\bibnamefont
  {McManus}}, \bibinfo {author} {\bibfnamefont {E.}~\bibnamefont {Berti}},
  \bibinfo {author} {\bibfnamefont {C.~F.~B.}\ \bibnamefont {Macedo}}, \bibinfo
  {author} {\bibfnamefont {M.}~\bibnamefont {Kimura}}, \bibinfo {author}
  {\bibfnamefont {A.}~\bibnamefont {Maselli}}, \ and\ \bibinfo {author}
  {\bibfnamefont {V.}~\bibnamefont {Cardoso}},\ }\href {\doibase
  10.1103/PhysRevD.100.044061} {\bibfield  {journal} {\bibinfo  {journal}
  {Phys. Rev.}\ }\textbf {\bibinfo {volume} {D100}},\ \bibinfo {pages} {044061}
  (\bibinfo {year} {2019})},\ \Eprint {http://arxiv.org/abs/1906.05155}
  {arXiv:1906.05155 [gr-qc]} \BibitemShut {NoStop}%
%%CITATION = ARXIV:1906.05155;%%
\bibitem [{\citenamefont {Cardoso}\ and\ \citenamefont
  {Pani}(2019)}]{Cardoso:2019rvt}%
  \BibitemOpen
  \bibfield  {author} {\bibinfo {author} {\bibfnamefont {V.}~\bibnamefont
  {Cardoso}}\ and\ \bibinfo {author} {\bibfnamefont {P.}~\bibnamefont {Pani}},\
  }\href {\doibase 10.1007/s41114-019-0020-4} {\bibfield  {journal} {\bibinfo
  {journal} {Living Rev. Rel.}\ }\textbf {\bibinfo {volume} {22}},\ \bibinfo
  {pages} {4} (\bibinfo {year} {2019})},\ \Eprint
  {http://arxiv.org/abs/1904.05363} {arXiv:1904.05363 [gr-qc]} \BibitemShut
  {NoStop}%
%%CITATION = ARXIV:1904.05363;%%
\bibitem [{\citenamefont {Chirenti}\ and\ \citenamefont
  {Rezzolla}(2016)}]{Chirenti:2016hzd}%
  \BibitemOpen
  \bibfield  {author} {\bibinfo {author} {\bibfnamefont {C.}~\bibnamefont
  {Chirenti}}\ and\ \bibinfo {author} {\bibfnamefont {L.}~\bibnamefont
  {Rezzolla}},\ }\href {\doibase 10.1103/PhysRevD.94.084016} {\bibfield
  {journal} {\bibinfo  {journal} {Phys. Rev.}\ }\textbf {\bibinfo {volume}
  {D94}},\ \bibinfo {pages} {084016} (\bibinfo {year} {2016})},\ \Eprint
  {http://arxiv.org/abs/1602.08759} {arXiv:1602.08759 [gr-qc]} \BibitemShut
  {NoStop}%
%%CITATION = ARXIV:1602.08759;%%
\bibitem [{\citenamefont {Abbott}\ \emph
  {et~al.}(2016{\natexlab{b}})\citenamefont {Abbott} \emph
  {et~al.}}]{ligo-gw150914}%
  \BibitemOpen
  \bibfield  {author} {\bibinfo {author} {\bibfnamefont {B.~P.}\ \bibnamefont
  {Abbott}} \emph {et~al.} (\bibinfo {collaboration} {LIGO Scientific,
  Virgo}),\ }\href {\doibase 10.1103/PhysRevLett.116.061102} {\bibfield
  {journal} {\bibinfo  {journal} {Phys. Rev. Lett.}\ }\textbf {\bibinfo
  {volume} {116}},\ \bibinfo {pages} {061102} (\bibinfo {year}
  {2016}{\natexlab{b}})},\ \Eprint {http://arxiv.org/abs/1602.03837}
  {arXiv:1602.03837 [gr-qc]} \BibitemShut {NoStop}%
%%CITATION = ARXIV:1602.03837;%%
\bibitem [{\citenamefont {Cotesta}\ \emph {et~al.}(2018)\citenamefont
  {Cotesta}, \citenamefont {Buonanno}, \citenamefont {Bohé}, \citenamefont
  {Taracchini}, \citenamefont {Hinder},\ and\ \citenamefont
  {Ossokine}}]{Cotesta-2018oct}%
  \BibitemOpen
  \bibfield  {author} {\bibinfo {author} {\bibfnamefont {R.}~\bibnamefont
  {Cotesta}}, \bibinfo {author} {\bibfnamefont {A.}~\bibnamefont {Buonanno}},
  \bibinfo {author} {\bibfnamefont {A.}~\bibnamefont {Bohé}}, \bibinfo
  {author} {\bibfnamefont {A.}~\bibnamefont {Taracchini}}, \bibinfo {author}
  {\bibfnamefont {I.}~\bibnamefont {Hinder}}, \ and\ \bibinfo {author}
  {\bibfnamefont {S.}~\bibnamefont {Ossokine}},\ }\href {\doibase
  10.1103/PhysRevD.98.084028} {\bibfield  {journal} {\bibinfo  {journal} {Phys.
  Rev.}\ }\textbf {\bibinfo {volume} {D98}},\ \bibinfo {pages} {084028}
  (\bibinfo {year} {2018})},\ \Eprint {http://arxiv.org/abs/1803.10701}
  {arXiv:1803.10701 [gr-qc]} \BibitemShut {NoStop}%
%%CITATION = ARXIV:1803.10701;%%
\bibitem [{\citenamefont {Kamaretsos}\ \emph {et~al.}(2012)\citenamefont
  {Kamaretsos}, \citenamefont {Hannam}, \citenamefont {Husa},\ and\
  \citenamefont {Sathyaprakash}}]{Kamaretsos:2011um}%
  \BibitemOpen
  \bibfield  {author} {\bibinfo {author} {\bibfnamefont {I.}~\bibnamefont
  {Kamaretsos}}, \bibinfo {author} {\bibfnamefont {M.}~\bibnamefont {Hannam}},
  \bibinfo {author} {\bibfnamefont {S.}~\bibnamefont {Husa}}, \ and\ \bibinfo
  {author} {\bibfnamefont {B.~S.}\ \bibnamefont {Sathyaprakash}},\ }\href
  {\doibase 10.1103/PhysRevD.85.024018} {\bibfield  {journal} {\bibinfo
  {journal} {Phys. Rev.}\ }\textbf {\bibinfo {volume} {D85}},\ \bibinfo {pages}
  {024018} (\bibinfo {year} {2012})},\ \Eprint {http://arxiv.org/abs/1107.0854}
  {arXiv:1107.0854 [gr-qc]} \BibitemShut {NoStop}%
%%CITATION = ARXIV:1107.0854;%%
\bibitem [{\citenamefont {Kelly}\ and\ \citenamefont
  {Baker}(2013)}]{Kelly:2012nd}%
  \BibitemOpen
  \bibfield  {author} {\bibinfo {author} {\bibfnamefont {B.~J.}\ \bibnamefont
  {Kelly}}\ and\ \bibinfo {author} {\bibfnamefont {J.~G.}\ \bibnamefont
  {Baker}},\ }\href {\doibase 10.1103/PhysRevD.87.084004} {\bibfield  {journal}
  {\bibinfo  {journal} {Phys. Rev.}\ }\textbf {\bibinfo {volume} {D87}},\
  \bibinfo {pages} {084004} (\bibinfo {year} {2013})},\ \Eprint
  {http://arxiv.org/abs/1212.5553} {arXiv:1212.5553 [gr-qc]} \BibitemShut
  {NoStop}%
%%CITATION = ARXIV:1212.5553;%%
\bibitem [{\citenamefont {Shi}\ \emph {et~al.}(2019)\citenamefont {Shi},
  \citenamefont {Bao}, \citenamefont {Wang}, \citenamefont {Zhang},
  \citenamefont {Hu}, \citenamefont {Sesana}, \citenamefont {Barausse},
  \citenamefont {Mei},\ and\ \citenamefont {Luo}}]{Shi:2019hqa}%
  \BibitemOpen
  \bibfield  {author} {\bibinfo {author} {\bibfnamefont {C.}~\bibnamefont
  {Shi}}, \bibinfo {author} {\bibfnamefont {J.}~\bibnamefont {Bao}}, \bibinfo
  {author} {\bibfnamefont {H.}~\bibnamefont {Wang}}, \bibinfo {author}
  {\bibfnamefont {J.-d.}\ \bibnamefont {Zhang}}, \bibinfo {author}
  {\bibfnamefont {Y.}~\bibnamefont {Hu}}, \bibinfo {author} {\bibfnamefont
  {A.}~\bibnamefont {Sesana}}, \bibinfo {author} {\bibfnamefont
  {E.}~\bibnamefont {Barausse}}, \bibinfo {author} {\bibfnamefont
  {J.}~\bibnamefont {Mei}}, \ and\ \bibinfo {author} {\bibfnamefont
  {J.}~\bibnamefont {Luo}},\ }\href {\doibase 10.1103/PhysRevD.100.044036}
  {\bibfield  {journal} {\bibinfo  {journal} {Phys. Rev.}\ }\textbf {\bibinfo
  {volume} {D100}},\ \bibinfo {pages} {044036} (\bibinfo {year} {2019})},\
  \Eprint {http://arxiv.org/abs/1902.08922} {arXiv:1902.08922 [gr-qc]}
  \BibitemShut {NoStop}%
%%CITATION = ARXIV:1902.08922;%%
\bibitem [{\citenamefont {Thrane}\ \emph {et~al.}(2017)\citenamefont {Thrane},
  \citenamefont {Lasky},\ and\ \citenamefont {Levin}}]{thrane-2017nov}%
  \BibitemOpen
  \bibfield  {author} {\bibinfo {author} {\bibfnamefont {E.}~\bibnamefont
  {Thrane}}, \bibinfo {author} {\bibfnamefont {P.~D.}\ \bibnamefont {Lasky}}, \
  and\ \bibinfo {author} {\bibfnamefont {Y.}~\bibnamefont {Levin}},\ }\href
  {\doibase 10.1103/PhysRevD.96.102004} {\bibfield  {journal} {\bibinfo
  {journal} {Phys. Rev.}\ }\textbf {\bibinfo {volume} {D96}},\ \bibinfo {pages}
  {102004} (\bibinfo {year} {2017})},\ \Eprint
  {http://arxiv.org/abs/1706.05152} {arXiv:1706.05152 [gr-qc]} \BibitemShut
  {NoStop}%
%%CITATION = ARXIV:1706.05152;%%
\bibitem [{\citenamefont {Maselli}\ \emph {et~al.}(2020)\citenamefont
  {Maselli}, \citenamefont {Pani}, \citenamefont {Gualtieri},\ and\
  \citenamefont {Berti}}]{Maselli:2019mjd}%
  \BibitemOpen
  \bibfield  {author} {\bibinfo {author} {\bibfnamefont {A.}~\bibnamefont
  {Maselli}}, \bibinfo {author} {\bibfnamefont {P.}~\bibnamefont {Pani}},
  \bibinfo {author} {\bibfnamefont {L.}~\bibnamefont {Gualtieri}}, \ and\
  \bibinfo {author} {\bibfnamefont {E.}~\bibnamefont {Berti}},\ }\href
  {\doibase 10.1103/PhysRevD.101.024043} {\bibfield  {journal} {\bibinfo
  {journal} {Phys. Rev.}\ }\textbf {\bibinfo {volume} {D101}},\ \bibinfo
  {pages} {024043} (\bibinfo {year} {2020})},\ \Eprint
  {http://arxiv.org/abs/1910.12893} {arXiv:1910.12893 [gr-qc]} \BibitemShut
  {NoStop}%
%%CITATION = ARXIV:1910.12893;%%
\bibitem [{\citenamefont {Buonanno}\ \emph {et~al.}(2007)\citenamefont
  {Buonanno}, \citenamefont {Cook},\ and\ \citenamefont
  {Pretorius}}]{Buonanno:2006ui}%
  \BibitemOpen
  \bibfield  {author} {\bibinfo {author} {\bibfnamefont {A.}~\bibnamefont
  {Buonanno}}, \bibinfo {author} {\bibfnamefont {G.~B.}\ \bibnamefont {Cook}},
  \ and\ \bibinfo {author} {\bibfnamefont {F.}~\bibnamefont {Pretorius}},\
  }\href {\doibase 10.1103/PhysRevD.75.124018} {\bibfield  {journal} {\bibinfo
  {journal} {Phys. Rev.}\ }\textbf {\bibinfo {volume} {D75}},\ \bibinfo {pages}
  {124018} (\bibinfo {year} {2007})},\ \Eprint
  {http://arxiv.org/abs/gr-qc/0610122} {arXiv:gr-qc/0610122 [gr-qc]}
  \BibitemShut {NoStop}%
%%CITATION = GR-QC/0610122;%%
\bibitem [{\citenamefont {London}\ \emph {et~al.}(2014)\citenamefont {London},
  \citenamefont {Shoemaker},\ and\ \citenamefont {Healy}}]{London:2014cma}%
  \BibitemOpen
  \bibfield  {author} {\bibinfo {author} {\bibfnamefont {L.}~\bibnamefont
  {London}}, \bibinfo {author} {\bibfnamefont {D.}~\bibnamefont {Shoemaker}}, \
  and\ \bibinfo {author} {\bibfnamefont {J.}~\bibnamefont {Healy}},\ }\href
  {\doibase 10.1103/PhysRevD.90.124032, 10.1103/PhysRevD.94.069902} {\bibfield
  {journal} {\bibinfo  {journal} {Phys. Rev.}\ }\textbf {\bibinfo {volume}
  {D90}},\ \bibinfo {pages} {124032} (\bibinfo {year} {2014})},\ \bibinfo
  {note} {[Erratum: Phys. Rev.D94,no.6,069902(2016)]},\ \Eprint
  {http://arxiv.org/abs/1404.3197} {arXiv:1404.3197 [gr-qc]} \BibitemShut
  {NoStop}%
%%CITATION = ARXIV:1404.3197;%%
\bibitem [{\citenamefont {Giesler}\ \emph {et~al.}(2019)\citenamefont
  {Giesler}, \citenamefont {Isi}, \citenamefont {Scheel},\ and\ \citenamefont
  {Teukolsky}}]{Giesler:2019uxc}%
  \BibitemOpen
  \bibfield  {author} {\bibinfo {author} {\bibfnamefont {M.}~\bibnamefont
  {Giesler}}, \bibinfo {author} {\bibfnamefont {M.}~\bibnamefont {Isi}},
  \bibinfo {author} {\bibfnamefont {M.}~\bibnamefont {Scheel}}, \ and\ \bibinfo
  {author} {\bibfnamefont {S.}~\bibnamefont {Teukolsky}},\ }\href {\doibase
  10.1103/PhysRevX.9.041060} {\bibfield  {journal} {\bibinfo  {journal} {Phys.
  Rev.}\ }\textbf {\bibinfo {volume} {X9}},\ \bibinfo {pages} {041060}
  (\bibinfo {year} {2019})},\ \Eprint {http://arxiv.org/abs/1903.08284}
  {arXiv:1903.08284 [gr-qc]} \BibitemShut {NoStop}%
%%CITATION = ARXIV:1903.08284;%%
\bibitem [{\citenamefont {Carullo}\ \emph {et~al.}(2019)\citenamefont
  {Carullo}, \citenamefont {Del~Pozzo},\ and\ \citenamefont
  {Veitch}}]{carullo-2019jun}%
  \BibitemOpen
  \bibfield  {author} {\bibinfo {author} {\bibfnamefont {G.}~\bibnamefont
  {Carullo}}, \bibinfo {author} {\bibfnamefont {W.}~\bibnamefont {Del~Pozzo}},
  \ and\ \bibinfo {author} {\bibfnamefont {J.}~\bibnamefont {Veitch}},\ }\href
  {\doibase 10.1103/PhysRevD.99.123029} {\bibfield  {journal} {\bibinfo
  {journal} {Phys. Rev.}\ }\textbf {\bibinfo {volume} {D99}},\ \bibinfo {pages}
  {123029} (\bibinfo {year} {2019})},\ \Eprint
  {http://arxiv.org/abs/1902.07527} {arXiv:1902.07527 [gr-qc]} \BibitemShut
  {NoStop}%
%%CITATION = ARXIV:1902.07527;%%
\bibitem [{\citenamefont {Abbott}\ \emph
  {et~al.}(2016{\natexlab{c}})\citenamefont {Abbott} \emph
  {et~al.}}]{ligo-propsgw150914}%
  \BibitemOpen
  \bibfield  {author} {\bibinfo {author} {\bibfnamefont {B.~P.}\ \bibnamefont
  {Abbott}} \emph {et~al.} (\bibinfo {collaboration} {LIGO Scientific,
  Virgo}),\ }\href {\doibase 10.1103/PhysRevLett.116.241102} {\bibfield
  {journal} {\bibinfo  {journal} {Phys. Rev. Lett.}\ }\textbf {\bibinfo
  {volume} {116}},\ \bibinfo {pages} {241102} (\bibinfo {year}
  {2016}{\natexlab{c}})},\ \Eprint {http://arxiv.org/abs/1602.03840}
  {arXiv:1602.03840 [gr-qc]} \BibitemShut {NoStop}%
%%CITATION = ARXIV:1602.03840;%%
\bibitem [{\citenamefont {Isi}\ \emph {et~al.}(2019)\citenamefont {Isi},
  \citenamefont {Giesler}, \citenamefont {Farr}, \citenamefont {Scheel},\ and\
  \citenamefont {Teukolsky}}]{Isi:2019aib}%
  \BibitemOpen
  \bibfield  {author} {\bibinfo {author} {\bibfnamefont {M.}~\bibnamefont
  {Isi}}, \bibinfo {author} {\bibfnamefont {M.}~\bibnamefont {Giesler}},
  \bibinfo {author} {\bibfnamefont {W.~M.}\ \bibnamefont {Farr}}, \bibinfo
  {author} {\bibfnamefont {M.~A.}\ \bibnamefont {Scheel}}, \ and\ \bibinfo
  {author} {\bibfnamefont {S.~A.}\ \bibnamefont {Teukolsky}},\ }\href {\doibase
  10.1103/PhysRevLett.123.111102} {\bibfield  {journal} {\bibinfo  {journal}
  {Phys. Rev. Lett.}\ }\textbf {\bibinfo {volume} {123}},\ \bibinfo {pages}
  {111102} (\bibinfo {year} {2019})},\ \Eprint
  {http://arxiv.org/abs/1905.00869} {arXiv:1905.00869 [gr-qc]} \BibitemShut
  {NoStop}%
%%CITATION = ARXIV:1905.00869;%%
\bibitem [{\citenamefont {Bhagwat}\ \emph {et~al.}(2020)\citenamefont
  {Bhagwat}, \citenamefont {Forteza}, \citenamefont {Pani},\ and\ \citenamefont
  {Ferrari}}]{Bhagwat:2019dtm}%
  \BibitemOpen
  \bibfield  {author} {\bibinfo {author} {\bibfnamefont {S.}~\bibnamefont
  {Bhagwat}}, \bibinfo {author} {\bibfnamefont {X.~J.}\ \bibnamefont
  {Forteza}}, \bibinfo {author} {\bibfnamefont {P.}~\bibnamefont {Pani}}, \
  and\ \bibinfo {author} {\bibfnamefont {V.}~\bibnamefont {Ferrari}},\ }\href
  {\doibase 10.1103/PhysRevD.101.044033} {\bibfield  {journal} {\bibinfo
  {journal} {Phys. Rev.}\ }\textbf {\bibinfo {volume} {D101}},\ \bibinfo
  {pages} {044033} (\bibinfo {year} {2020})},\ \Eprint
  {http://arxiv.org/abs/1910.08708} {arXiv:1910.08708 [gr-qc]} \BibitemShut
  {NoStop}%
%%CITATION = ARXIV:1910.08708;%%
\bibitem [{\citenamefont {Dorband}\ \emph {et~al.}(2006)\citenamefont
  {Dorband}, \citenamefont {Berti}, \citenamefont {Diener}, \citenamefont
  {Schnetter},\ and\ \citenamefont {Tiglio}}]{Dorband-2006oct}%
  \BibitemOpen
  \bibfield  {author} {\bibinfo {author} {\bibfnamefont {E.~N.}\ \bibnamefont
  {Dorband}}, \bibinfo {author} {\bibfnamefont {E.}~\bibnamefont {Berti}},
  \bibinfo {author} {\bibfnamefont {P.}~\bibnamefont {Diener}}, \bibinfo
  {author} {\bibfnamefont {E.}~\bibnamefont {Schnetter}}, \ and\ \bibinfo
  {author} {\bibfnamefont {M.}~\bibnamefont {Tiglio}},\ }\href {\doibase
  10.1103/PhysRevD.74.084028} {\bibfield  {journal} {\bibinfo  {journal} {Phys.
  Rev.}\ }\textbf {\bibinfo {volume} {D74}},\ \bibinfo {pages} {084028}
  (\bibinfo {year} {2006})},\ \Eprint {http://arxiv.org/abs/gr-qc/0608091}
  {arXiv:gr-qc/0608091 [gr-qc]} \BibitemShut {NoStop}%
%%CITATION = GR-QC/0608091;%%
\bibitem [{\citenamefont {Berti}\ \emph {et~al.}(2007)\citenamefont {Berti},
  \citenamefont {Cardoso}, \citenamefont {Gonzalez}, \citenamefont {Sperhake},
  \citenamefont {Hannam}, \citenamefont {Husa},\ and\ \citenamefont
  {Bruegmann}}]{berti-2007sep}%
  \BibitemOpen
  \bibfield  {author} {\bibinfo {author} {\bibfnamefont {E.}~\bibnamefont
  {Berti}}, \bibinfo {author} {\bibfnamefont {V.}~\bibnamefont {Cardoso}},
  \bibinfo {author} {\bibfnamefont {J.~A.}\ \bibnamefont {Gonzalez}}, \bibinfo
  {author} {\bibfnamefont {U.}~\bibnamefont {Sperhake}}, \bibinfo {author}
  {\bibfnamefont {M.}~\bibnamefont {Hannam}}, \bibinfo {author} {\bibfnamefont
  {S.}~\bibnamefont {Husa}}, \ and\ \bibinfo {author} {\bibfnamefont
  {B.}~\bibnamefont {Bruegmann}},\ }\href {\doibase 10.1103/PhysRevD.76.064034}
  {\bibfield  {journal} {\bibinfo  {journal} {Phys. Rev.}\ }\textbf {\bibinfo
  {volume} {D76}},\ \bibinfo {pages} {064034} (\bibinfo {year} {2007})},\
  \Eprint {http://arxiv.org/abs/gr-qc/0703053} {arXiv:gr-qc/0703053 [GR-QC]}
  \BibitemShut {NoStop}%
%%CITATION = GR-QC/0703053;%%
\bibitem [{\citenamefont {Carullo}\ \emph {et~al.}(2018)\citenamefont {Carullo}
  \emph {et~al.}}]{Carullo-2018nov}%
  \BibitemOpen
  \bibfield  {author} {\bibinfo {author} {\bibfnamefont {G.}~\bibnamefont
  {Carullo}} \emph {et~al.},\ }\href {\doibase 10.1103/PhysRevD.98.104020}
  {\bibfield  {journal} {\bibinfo  {journal} {Phys. Rev.}\ }\textbf {\bibinfo
  {volume} {D98}},\ \bibinfo {pages} {104020} (\bibinfo {year} {2018})},\
  \Eprint {http://arxiv.org/abs/1805.04760} {arXiv:1805.04760 [gr-qc]}
  \BibitemShut {NoStop}%
%%CITATION = ARXIV:1805.04760;%%
\bibitem [{\citenamefont {Boyle}\ \emph {et~al.}(2019)\citenamefont {Boyle}
  \emph {et~al.}}]{Boyle:2019kee}%
  \BibitemOpen
  \bibfield  {author} {\bibinfo {author} {\bibfnamefont {M.}~\bibnamefont
  {Boyle}} \emph {et~al.},\ }\href {\doibase 10.1088/1361-6382/ab34e2}
  {\bibfield  {journal} {\bibinfo  {journal} {Class. Quant. Grav.}\ }\textbf
  {\bibinfo {volume} {36}},\ \bibinfo {pages} {195006} (\bibinfo {year}
  {2019})},\ \Eprint {http://arxiv.org/abs/1904.04831} {arXiv:1904.04831
  [gr-qc]} \BibitemShut {NoStop}%
%%CITATION = ARXIV:1904.04831;%%
\bibitem [{SXS()}]{SXS-catalog}%
  \BibitemOpen
  \href@noop {} {}\bibinfo {howpublished}
  {\url{https://data.black-holes.org/waveforms/index.html}}\BibitemShut
  {NoStop}%
\bibitem [{\citenamefont {Ferguson}\ \emph {et~al.}(2019)\citenamefont
  {Ferguson}, \citenamefont {Ghonge}, \citenamefont {Clark}, \citenamefont
  {Calderon~Bustillo}, \citenamefont {Laguna}, \citenamefont {Shoemaker},\ and\
  \citenamefont {Calderon~Bustillo}}]{Ferguson-2019}%
  \BibitemOpen
  \bibfield  {author} {\bibinfo {author} {\bibfnamefont {D.}~\bibnamefont
  {Ferguson}}, \bibinfo {author} {\bibfnamefont {S.}~\bibnamefont {Ghonge}},
  \bibinfo {author} {\bibfnamefont {J.~A.}\ \bibnamefont {Clark}}, \bibinfo
  {author} {\bibfnamefont {J.}~\bibnamefont {Calderon~Bustillo}}, \bibinfo
  {author} {\bibfnamefont {P.}~\bibnamefont {Laguna}}, \bibinfo {author}
  {\bibfnamefont {D.}~\bibnamefont {Shoemaker}}, \ and\ \bibinfo {author}
  {\bibfnamefont {J.}~\bibnamefont {Calderon~Bustillo}},\ }\href {\doibase
  10.1103/PhysRevLett.123.151101} {\bibfield  {journal} {\bibinfo  {journal}
  {Phys. Rev. Lett.}\ }\textbf {\bibinfo {volume} {123}},\ \bibinfo {pages}
  {151101} (\bibinfo {year} {2019})},\ \Eprint
  {http://arxiv.org/abs/1905.03756} {arXiv:1905.03756 [gr-qc]} \BibitemShut
  {NoStop}%
%%CITATION = ARXIV:1905.03756;%%
\bibitem [{Ber()}]{Berti-ringdown}%
  \BibitemOpen
  \href@noop {} {}\bibinfo {howpublished}
  {\url{https://pages.jh.edu/~eberti2/ringdown/}}\BibitemShut {NoStop}%
\bibitem [{\citenamefont {Berti}\ and\ \citenamefont
  {Cardoso}(2006)}]{Berti:2006wq}%
  \BibitemOpen
  \bibfield  {author} {\bibinfo {author} {\bibfnamefont {E.}~\bibnamefont
  {Berti}}\ and\ \bibinfo {author} {\bibfnamefont {V.}~\bibnamefont
  {Cardoso}},\ }\href {\doibase 10.1103/PhysRevD.74.104020} {\bibfield
  {journal} {\bibinfo  {journal} {Phys. Rev.}\ }\textbf {\bibinfo {volume}
  {D74}},\ \bibinfo {pages} {104020} (\bibinfo {year} {2006})},\ \Eprint
  {http://arxiv.org/abs/gr-qc/0605118} {arXiv:gr-qc/0605118 [gr-qc]}
  \BibitemShut {NoStop}%
%%CITATION = GR-QC/0605118;%%
\bibitem [{\citenamefont {Chirenti}\ and\ \citenamefont
  {Rezzolla}(2007)}]{Chirenti_2007}%
  \BibitemOpen
  \bibfield  {author} {\bibinfo {author} {\bibfnamefont {C.~B. M.~H.}\
  \bibnamefont {Chirenti}}\ and\ \bibinfo {author} {\bibfnamefont
  {L.}~\bibnamefont {Rezzolla}},\ }\href {\doibase 10.1088/0264-9381/24/16/013}
  {\bibfield  {journal} {\bibinfo  {journal} {Classical and Quantum Gravity}\
  }\textbf {\bibinfo {volume} {24}},\ \bibinfo {pages} {4191–4206} (\bibinfo
  {year} {2007})}\BibitemShut {NoStop}%
\bibitem [{\citenamefont {Nollert}(1999{\natexlab{a}})}]{8marcel-grossman}%
  \BibitemOpen
  \bibfield  {author} {\bibinfo {author} {\bibfnamefont {H.-P.}\ \bibnamefont
  {Nollert}},\ }\href@noop {} {\emph {\bibinfo {title} {The Eighth Marcel
  Grossmann Meeting}}}\ (\bibinfo  {publisher} {World Scientific Pub Co Inc},\
  \bibinfo {year} {1999})\BibitemShut {NoStop}%
\bibitem [{\citenamefont {Nollert}(1999{\natexlab{b}})}]{Nollert_1999}%
  \BibitemOpen
  \bibfield  {author} {\bibinfo {author} {\bibfnamefont {H.-P.}\ \bibnamefont
  {Nollert}},\ }\href {\doibase 10.1088/0264-9381/16/12/201} {\bibfield
  {journal} {\bibinfo  {journal} {Class. Quant. Grav.}\ }\textbf {\bibinfo
  {volume} {16}},\ \bibinfo {pages} {R159} (\bibinfo {year}
  {1999}{\natexlab{b}})}\BibitemShut {NoStop}%
%%CITATION = CQGRD,16,R159;%%
\bibitem [{GW1()}]{GW150914-data}%
  \BibitemOpen
  \href@noop {} {}\bibinfo {howpublished}
  {\url{https://www.gw-openscience.org/events/GW150914/}}\BibitemShut {NoStop}%
\bibitem [{\citenamefont {Moore}\ \emph {et~al.}(2015)\citenamefont {Moore},
  \citenamefont {Cole},\ and\ \citenamefont {Berry}}]{Moore:2014lga}%
  \BibitemOpen
  \bibfield  {author} {\bibinfo {author} {\bibfnamefont {C.~J.}\ \bibnamefont
  {Moore}}, \bibinfo {author} {\bibfnamefont {R.~H.}\ \bibnamefont {Cole}}, \
  and\ \bibinfo {author} {\bibfnamefont {C.~P.~L.}\ \bibnamefont {Berry}},\
  }\href {\doibase 10.1088/0264-9381/32/1/015014} {\bibfield  {journal}
  {\bibinfo  {journal} {Class. Quant. Grav.}\ }\textbf {\bibinfo {volume}
  {32}},\ \bibinfo {pages} {015014} (\bibinfo {year} {2015})},\ \Eprint
  {http://arxiv.org/abs/1408.0740} {arXiv:1408.0740 [gr-qc]} \BibitemShut
  {NoStop}%
%%CITATION = ARXIV:1408.0740;%%
\bibitem [{\citenamefont {Finn}(1992)}]{Finn:1992wt}%
  \BibitemOpen
  \bibfield  {author} {\bibinfo {author} {\bibfnamefont {L.~S.}\ \bibnamefont
  {Finn}},\ }\href {\doibase 10.1103/PhysRevD.46.5236} {\bibfield  {journal}
  {\bibinfo  {journal} {Phys. Rev.}\ }\textbf {\bibinfo {volume} {D46}},\
  \bibinfo {pages} {5236} (\bibinfo {year} {1992})},\ \Eprint
  {http://arxiv.org/abs/gr-qc/9209010} {arXiv:gr-qc/9209010 [gr-qc]}
  \BibitemShut {NoStop}%
%%CITATION = GR-QC/9209010;%%
\bibitem [{Des()}]{Design-sensitivity}%
  \BibitemOpen
  \href@noop {} {}\bibinfo {howpublished}
  {\url{https://dcc.ligo.org/LIGO-T1800044/public}}\BibitemShut {NoStop}%
\bibitem [{\citenamefont {Maggiore}(2008)}]{maggiore2008gravitational}%
  \BibitemOpen
  \bibfield  {author} {\bibinfo {author} {\bibfnamefont {M.}~\bibnamefont
  {Maggiore}},\ }\href@noop {} {\emph {\bibinfo {title} {Gravitational Waves:
  Volume 1: Theory and Experiments}}},\ Gravitational Waves\ (\bibinfo
  {publisher} {Oxford University Press},\ \bibinfo {year} {2008})\BibitemShut
  {NoStop}%
\bibitem [{\citenamefont {{Berti}}\ \emph {et~al.}(2006)\citenamefont
  {{Berti}}, \citenamefont {{Cardoso}},\ and\ \citenamefont
  {{Casals}}}]{berti.cardoso.casals}%
  \BibitemOpen
  \bibfield  {author} {\bibinfo {author} {\bibfnamefont {E.}~\bibnamefont
  {{Berti}}}, \bibinfo {author} {\bibfnamefont {V.}~\bibnamefont {{Cardoso}}},
  \ and\ \bibinfo {author} {\bibfnamefont {M.}~\bibnamefont {{Casals}}},\
  }\href {\doibase 10.1103/PhysRevD.73.024013} {\bibfield  {journal} {\bibinfo
  {journal} {\prd}\ }\textbf {\bibinfo {volume} {73}},\ \bibinfo {eid} {024013}
  (\bibinfo {year} {2006})},\ \Eprint {http://arxiv.org/abs/gr-qc/0511111}
  {arXiv:gr-qc/0511111 [gr-qc]} \BibitemShut {NoStop}%
\bibitem [{\citenamefont {Venumadhav}\ \emph {et~al.}(2019)\citenamefont
  {Venumadhav}, \citenamefont {Zackay}, \citenamefont {Roulet}, \citenamefont
  {Dai},\ and\ \citenamefont {Zaldarriaga}}]{venumadhav2019new}%
  \BibitemOpen
  \bibfield  {author} {\bibinfo {author} {\bibfnamefont {T.}~\bibnamefont
  {Venumadhav}}, \bibinfo {author} {\bibfnamefont {B.}~\bibnamefont {Zackay}},
  \bibinfo {author} {\bibfnamefont {J.}~\bibnamefont {Roulet}}, \bibinfo
  {author} {\bibfnamefont {L.}~\bibnamefont {Dai}}, \ and\ \bibinfo {author}
  {\bibfnamefont {M.}~\bibnamefont {Zaldarriaga}},\ }\href@noop {} {\
  (\bibinfo {year} {2019})},\ \Eprint {http://arxiv.org/abs/1904.07214}
  {arXiv:1904.07214 [astro-ph.HE]} \BibitemShut {NoStop}%
\bibitem [{\citenamefont {Zackay}\ \emph {et~al.}(2019)\citenamefont {Zackay},
  \citenamefont {Dai}, \citenamefont {Venumadhav}, \citenamefont {Roulet},\
  and\ \citenamefont {Zaldarriaga}}]{zackay2019detecting}%
  \BibitemOpen
  \bibfield  {author} {\bibinfo {author} {\bibfnamefont {B.}~\bibnamefont
  {Zackay}}, \bibinfo {author} {\bibfnamefont {L.}~\bibnamefont {Dai}},
  \bibinfo {author} {\bibfnamefont {T.}~\bibnamefont {Venumadhav}}, \bibinfo
  {author} {\bibfnamefont {J.}~\bibnamefont {Roulet}}, \ and\ \bibinfo {author}
  {\bibfnamefont {M.}~\bibnamefont {Zaldarriaga}},\ }\href@noop {} {\
  (\bibinfo {year} {2019})},\ \Eprint {http://arxiv.org/abs/1910.09528}
  {arXiv:1910.09528 [astro-ph.HE]} \BibitemShut {NoStop}%
\bibitem [{\citenamefont {Chatziioannou}\ \emph {et~al.}(2019)\citenamefont
  {Chatziioannou} \emph {et~al.}}]{Chatziioannou:2019dsz}%
  \BibitemOpen
  \bibfield  {author} {\bibinfo {author} {\bibfnamefont {K.}~\bibnamefont
  {Chatziioannou}} \emph {et~al.},\ }\href {\doibase
  10.1103/PhysRevD.100.104015} {\bibfield  {journal} {\bibinfo  {journal}
  {Phys. Rev.}\ }\textbf {\bibinfo {volume} {D100}},\ \bibinfo {pages} {104015}
  (\bibinfo {year} {2019})},\ \Eprint {http://arxiv.org/abs/1903.06742}
  {arXiv:1903.06742 [gr-qc]} \BibitemShut {NoStop}%
%%CITATION = ARXIV:1903.06742;%%
\bibitem [{\citenamefont {Yang}\ \emph {et~al.}(2017)\citenamefont {Yang},
  \citenamefont {Yagi}, \citenamefont {Blackman}, \citenamefont {Lehner},
  \citenamefont {Paschalidis}, \citenamefont {Pretorius},\ and\ \citenamefont
  {Yunes}}]{Yang_2017}%
  \BibitemOpen
  \bibfield  {author} {\bibinfo {author} {\bibfnamefont {H.}~\bibnamefont
  {Yang}}, \bibinfo {author} {\bibfnamefont {K.}~\bibnamefont {Yagi}}, \bibinfo
  {author} {\bibfnamefont {J.}~\bibnamefont {Blackman}}, \bibinfo {author}
  {\bibfnamefont {L.}~\bibnamefont {Lehner}}, \bibinfo {author} {\bibfnamefont
  {V.}~\bibnamefont {Paschalidis}}, \bibinfo {author} {\bibfnamefont
  {F.}~\bibnamefont {Pretorius}}, \ and\ \bibinfo {author} {\bibfnamefont
  {N.}~\bibnamefont {Yunes}},\ }\href {\doibase 10.1103/physrevlett.118.161101}
  {\bibfield  {journal} {\bibinfo  {journal} {Physical Review Letters}\
  }\textbf {\bibinfo {volume} {118}} (\bibinfo {year} {2017}),\
  10.1103/physrevlett.118.161101}\BibitemShut {NoStop}%
\bibitem [{\citenamefont {Berti}\ \emph {et~al.}(2018)\citenamefont {Berti},
  \citenamefont {Yagi}, \citenamefont {Yang},\ and\ \citenamefont
  {Yunes}}]{Berti_2018}%
  \BibitemOpen
  \bibfield  {author} {\bibinfo {author} {\bibfnamefont {E.}~\bibnamefont
  {Berti}}, \bibinfo {author} {\bibfnamefont {K.}~\bibnamefont {Yagi}},
  \bibinfo {author} {\bibfnamefont {H.}~\bibnamefont {Yang}}, \ and\ \bibinfo
  {author} {\bibfnamefont {N.}~\bibnamefont {Yunes}},\ }\href {\doibase
  10.1007/s10714-018-2372-6} {\bibfield  {journal} {\bibinfo  {journal}
  {General Relativity and Gravitation}\ }\textbf {\bibinfo {volume} {50}}
  (\bibinfo {year} {2018}),\ 10.1007/s10714-018-2372-6}\BibitemShut {NoStop}%
\end{thebibliography}%

\end{document}